\documentstyle[]{mn}
\newif\ifAMStwofonts
\def\ApJ{ApJ}
\def\MNRAS{MNRAS}

\def\cstar{C\(^{\rm *}\)}
\def\ee #1 {\times 10^{#1}}
\def\ut #1 #2 { \, \rmn{#1}^{#2}}
\def\u #1 { \, \rmn{#1}}

\def\kms {\,\rmn{km\,s}^{-1}}

\def\percc {\,\rmn{cm}^{-3}}

\def\half{{\scriptstyle \frac{1}{2}}}

\let\grad=\nabla
\def\cross{\bmath{\times}}
\def\curl #1 {\grad \cross #1}
\def\div #1 {\grad \cdot #1}
\def\nh{{n_{\rm H}}}
\def\ddz #1{{d\over dz} \left( #1 \right)}
\def\nh{{n_{\rm H}}}

\def\v{\bmath{v}}

\def\vj{\bmath{v}_j}

\def\B{\bmath{B}}
\def\Bh{\bmath{\hat{B}}}
\def\Bp{\bmath{B_{\perp}}}
\def\E{\bmath{E}}
\def\J{\bmath{J}}

\ifoldfss
  \newcommand{\rmn}[1] {{\rm #1}}

  \ifCUPmtlplainloaded \else
    \NewTextAlphabet{textbfit} {cmbxti10} {}
    \NewTextAlphabet{textbfss} {cmssbx10} {}
    \NewMathAlphabet{mathbfit} {cmbxti10} {} 
    \NewMathAlphabet{mathbfss} {cmssbx10} {} 
  \fi
  \ifAMStwofonts
    \ifCUPmtlplainloaded \else
      \NewSymbolFont{upmath} {eurm10}
      \NewSymbolFont{AMSa} {msam10}
      \NewMathSymbol{\upi}     {0}{upmath}{19}
      \NewMathSymbol{\umu}     {0}{upmath}{16}
      \NewMathSymbol{\upartial}{0}{upmath}{40}
      \NewMathSymbol{\leqslant}{3}{AMSa}{36}
      \NewMathSymbol{\geqslant}{3}{AMSa}{3E}

    \fi
  \fi
\fi 
\ifnfssone
  \newmathalphabet{\mathit}
  \addtoversion{normal}{\mathit}{cmr}{m}{it}
  \addtoversion{bold}{\mathit}{cmr}{bx}{it}
  \newcommand{\rmn}[1] {\mathrm{#1}}

  \newmathalphabet{\mathbfit} 
  \addtoversion{normal}{\mathbfit}{cmr}{bx}{it}
  \addtoversion{bold}{\mathbfit}{cmr}{bx}{it}
  \newmathalphabet{\mathbfss} 
  \addtoversion{normal}{\mathbfss}{cmss}{bx}{n}
  \addtoversion{bold}{\mathbfss}{cmss}{bx}{n}
  \ifAMStwofonts
    \ifCUPmtlplainloaded \else
      \UseAMStwoboldmath
      \makeatletter
      \new@mathgroup\upmath@group
      \define@mathgroup\mv@normal\upmath@group{eur}{m}{n}
      \define@mathgroup\mv@bold\upmath@group{eur}{b}{n}
      \edef\UPM{\hexnumber\upmath@group}
      \new@mathgroup\amsa@group
      \define@mathgroup\mv@normal\amsa@group{msa}{m}{n}
      \define@mathgroup\mv@bold\amsa@group{msa}{m}{n}
      \edef\AMSa{\hexnumber\amsa@group}
      \makeatother
      \mathchardef\upi="0\UPM19
      \mathchardef\umu="0\UPM16
      \mathchardef\upartial="0\UPM40
      \mathchardef\leqslant="3\AMSa36
      \mathchardef\geqslant="3\AMSa3E
    \fi
  \fi
\fi 
\ifnfsstwo
  \newcommand{\rmn}[1] {\mathrm{#1}}

  \DeclareMathAlphabet{\mathbfit}{OT1}{cmr}{bx}{it}
  \SetMathAlphabet\mathbfit{bold}{OT1}{cmr}{bx}{it}
  \DeclareMathAlphabet{\mathbfss}{OT1}{cmss}{bx}{n}
  \SetMathAlphabet\mathbfss{bold}{OT1}{cmss}{bx}{n}
  \ifAMStwofonts
    \ifCUPmtlplainloaded \else
      \DeclareSymbolFont{UPM}{U}{eur}{m}{n}
      \SetSymbolFont{UPM}{bold}{U}{eur}{b}{n}
      \DeclareSymbolFont{AMSa}{U}{msa}{m}{n}
      \DeclareMathSymbol{\upi}{0}{UPM}{"19}
      \DeclareMathSymbol{\umu}{0}{UPM}{"16}
      \DeclareMathSymbol{\upartial}{0}{UPM}{"40}
      \DeclareMathSymbol{\leqslant}{3}{AMSa}{"36}
      \DeclareMathSymbol{\geqslant}{3}{AMSa}{"3E}
    \fi
  \fi
\fi 
\ifCUPmtlplainloaded \else
  \ifAMStwofonts \else 
    \def\upi{\pi}
    \def\umu{\mu}
    \def\upartial{\partial}
  \fi
\fi
\input epsf
\title[Grains and C-type shocks in molecular clouds]
{Dust grains and the structure of steady C-type magnetohydrodynamic shock
waves in molecular clouds}
\author[M. Wardle]
       {Mark Wardle \\ 
Special Research Centre for Theoretical Astrophysics, University
of Sydney, NSW 2006, Australia
}
\pagerange{\pageref{firstpage}--\pageref{lastpage}}
\pubyear{1998}
\begin{document}
\maketitle
\label{firstpage}
\begin{abstract}
I examine the role of dust grains in determining the structure of 
steady, cold, oblique C-type shocks in dense molecular gas.  Gas 
pressure, the inertia of the charged components, and changes in 
ionization are neglected.  The grain charge and rate coefficients for 
electron-neutral and grain-neutral elastic scattering are assumed 
constant at values appropriate to the shock interior.  An MRN size 
distribution is accounted for by estimating an effective 
grain abundance and Hall parameter for single-size grains.

A one-parameter family of intermediate shocks exists for each shock 
speed \( v_s \) between the intermediate signal speed \(v_A {\rm cos} 
\theta \) and \(\sqrt{2} v_A {\rm cot} \theta\), where \( v_A \) is the 
pre-shock Alfv\'en speed and \(\theta\) is the angle between the pre-shock 
magnetic field and the normal to the shock front.  In addition, there 
is a unique fast shock for each \( v_s > v_A \).

If the pre-shock density \( \nh \ga 10^5 \percc \) and the pre-shock 
magnetic field satisfies \( B(\mathrm{mG}) /\nh(10^5\percc ) \la 1 \) 
grains are partially decoupled from the magnetic field and the field 
and velocity components within fast shocks do not lie in the plane 
containing the pre-shock field and the shock normal.  The resulting 
shock structure is significantly thinner than in models that do not 
take this into account.  Existing models systematically underestimate 
the grain-neutral drift speed and the heating rate within the shock 
front.  At densities in excess of \( 10^8 \percc \) these effects may 
be reduced by the nearly-equal abundances of positive and negative 
grains.

\end{abstract}
\begin{keywords}
magnetohydrodynamics -- shock waves -- dust grains.
\end{keywords}

\section{Introduction}

The structure of shock waves in molecular clouds is determined by the 
coupling between the magnetic field and the weakly ionized pre-shock 
gas.  For shock speeds below 40--50 \( \kms \) (McKee, Chernoff \& Hollenbach 
1984) Lorentz forces within the shock front push the charged species 
through the neutrals, and the resulting collisions accelerate, 
compress, and heat the ambient gas.  This process is slow because the 
charged species are rare, so the gas is able to radiate away a 
significant fraction of the heat while still within the shock front 
(Mullan 1971; Draine 1980), primarily through the emission of 
radiation in molecular rotational and vibrational transitions.  Shocks 
of this nature are denoted `C-type' (Draine 1980), or 
`\cstar-type' if the gas is decelerated through a sonic point in the 
reference frame comoving with the shock front (Chernoff 1987, Roberge 
\& Draine 1990).  At higher shock speeds, the molecular coolants are 
dissociated, cooling can no longer keep pace with heating, and the gas 
pressure becomes dynamically significant.  A thin, viscous sub-shock 
forms within the front, and the shock is termed `J-type' (Draine 
1980).

C-type shocks efficiently convert the heat produced by collisions 
within the shock front into molecular line emission, and are 
responsible for much of the intense infrared H\( _2 \) and CO line 
emission observed towards the Orion-KL region (Draine \& Roberge 1982; 
Chernoff, McKee \& Hollenbach 1982; Smith \& Brand 1990; Smith, Brand 
\& Moorhouse 1991; Chrysostomou et al. 1997).  The warm molecular 
environment within C-shocks can drive molecular chemistry such as the 
conversion of atomic oxygen into OH and then into water (Draine, 
Roberge \& Dalgarno 1983; Kaufman \& Neufeld 1996a,b).  In addition, 
the large drift speeds can drive exothermic ion-neutral chemistry 
within the shock front (Flower, Pineau des F\^orets \& Hartquist 1985; 
Pineau des F\^orets, Flower \& Hartquist 1986; Draine \& Katz 
1986a,b).

Shock models have generally been forced to be `coplanar', that is, the 
drift velocities and magnetic field within the shock front are assumed 
to lie in the `shock plane', the plane containing the pre-shock 
magnetic field and the shock normal (see Fig. 1).  This holds for 
fast\footnote{`Fast' and `slow' have occasionally been used in the 
astrophysical literature to denote J-type and C-type shocks 
respectively.  In this paper I use `fast' and `slow' in the 
traditional MHD sense, referring to the fast and slow signal speeds.} 
shocks if the charged species are tied to the field lines, but charged 
grains drift obliquely to the shock plane if they become partially 
decoupled from the field lines by collisions with the neutrals (Draine 
1980).  However, for the sake of simplicity, the component of the 
grain drag force on the neutrals perpendicular to the shock plane has 
usually been supressed (e.g.  Draine et al. 1983; Wardle 
\& Draine 1987).  This approximation breaks down for high pre-shock 
densities, partly because the fractional ionization of the gas 
decreases, permitting the grains to make a larger fractional 
contribution to the drag on the neutral gas, but also because the 
grains become poorly coupled to the magnetic field and the grain drag 
vector is further tilted out of the shock plane.

This motivated calculations of the structure of grain-dominated 
C-shocks in which the vector components perpendicular to the shock 
plane were retained.  Pilipp, Hartquist \& Havnes (1990) showed that 
the thickness of perpendicular shocks was significantly changed for 
pre-shock densities in excess of \(10^7 \percc\).  In this case, the 
symmetries of the geometry constrain the magnetic field direction to 
be the same throughout the shock front, even though the drift 
velocities need not lie within the shock plane.

Oblique shocks introduce an important extra degree of freedom, in that 
\( \Bp \), the magnetic field component perpendicular to the shock 
normal, may rotate within the shock front before returning to the 
shock plane downstream.  Oblique shocks were studied by Pilipp \& 
Hartquist (1994), who found solutions for shock speeds below a few 
\(\kms\) in which \( \Bp \) rotates by 180\degr (in either sense) 
within the shock front.  Pilipp \& Hartquist were unable to find 
solutions for shock velocities in excess of a few \(\kms\), and 
suggested that steady shock solutions may not exist in this regime.  
This is disconcerting, as shocks with speeds in excess of 20 \(\kms\) 
are required to model the observed emission from a variety of sources.  
In addition, the development of a sound theoretical understanding of 
magnetohydrodynamics of dense, dusty gas is important for scenarios 
for the formation and collapse of molecular cloud cores and the 
formation of protostellar discs (e.g.  Nishi, Nakano \& Umebayashi 
1991; Ciolek \& Mouschovias 1993; Neufeld \& Hollenbach 1994; Li \& 
McKee 1996; see also the review by Hartquist, Pilipp \& Havnes 1997).

The behaviour of \( \Bp \) in the solutions obtained by Pilipp \& 
Hartquist is fundamentally different to the earlier coplanar oblique 
shock models (Wardle \& Draine 1987; Wardle 1991a; Smith 1992), in 
which the pre-shock and post-shock \( \Bp \) vectors are parallel 
rather than antiparallel.  The upstream and downstream states across 
magnetohydrodynamic (MHD) shocks are related by jump conditions that 
are independent of the detailed nature and behaviour of the fluid, and 
therefore of the actual shock structure.  In particular, the 
transverse magnetic field increases, changes sign, or decreases across 
the fast, intermediate, or slow shocks respectively (e.g.  Cowling 
1976; Kennel, Blandford \& Coppi 1989).  This implies that the 
solutions found by Pilipp \& Hartquist are C-type \emph{intermediate} 
shocks, whereas previously studied C-shock solutions are \emph{fast} 
shocks.  The jump conditions show that the downstream state for 
intermediate shocks is unphysical for shock speeds a little above the 
Alfv\'en speed, and this explains the breakdown of the solutions for 
mildly super-Alfv\'enic shocks.  Further, for a given shock speed, 
Pilipp \& Hartquist found a one-parameter family of shock solutions 
rather than the single shock solution found in previous studies, and 
noted that this has also been found to be the case for intermediate 
shocks in resistive MHD (Wu 1988a; Kennel, Blandford \& Wu 1990).

Having identified these solutions as intermediate shocks, there still 
remains the important issue of the existence of fast shock solutions.  
On grounds of physical continuity, one expects that adding grains to a 
coplanar, fast C-type shock should distort the structure out of the 
shock plane rather than do away with it entirely.  Indeed, there are 
hints that steady shock solutions may exist.  Following standard 
practice in C-shock modelling, Pilipp \& Hartquist derived a set of 
ODE's describing the shock structure, and integrated them from a 
perturbed upstream state towards the shock front.  There is, however, 
a difference in the number of free parameters describing the initial 
perturbed state.  In the coplanar case there is one free parameter -- 
the amplitude of the initial perturbation in the pre-shock fluid 
(Draine 1980; Wardle \& Draine 1987).  Physically, this specifies the 
point in the shock precursor at which the integration towards the 
shock front begins, so there is a single shock solution for a given 
shock speed, the fast shock.  Dropping the assumption of coplanarity 
introduces a second parameter, which can be regarded as the degree of 
rotation of the perturbation out of the shock plane at the initial 
point.  The freedom introduced by the new parameter potentially allows 
a family of solutions to exist (the intermediate shocks), and Pilipp 
\& Hartquist found that there is a critical value at which the sense 
of rotation of \( \Bp \) across the intermediate shock solutions 
changes sign.  This suggests that the fast shock solution corresponds 
to this critical value, but that the solution cannot be found by 
integration from the perturbed upstream state because of finite 
numerical precision.

This paper demonstrates that fast non-coplanar C-shock do exist.  As 
the issues addressed here are fundamental in nature, I keep the 
formulation of the multifluid shock problem, described in the next 
section, as simple as possible.  In particular, I assume that the 
molecular gas consists of four distinct species: neutrals, positive 
ions, electrons or PAHs, and negatively charged grains.  Gas pressure 
is neglected, and ionization balance and chemistry are also ignored.  
These simplifications allow all of the physical quantities in the 
shock to be expressed in terms of the components, \(B_x\) and \(B_y\), 
of \( \Bp \).  Thus a shock solution can be represented by a plot of 
\(B_y(z)\) {\it vs.} \(B_x(z)\), where \( z \) is the coordinate along 
the shock normal.  This two-dimensional phase space, explored in 
\S\ref{sec-analysis}, is a powerful tool for examining the set of 
shock solutions given the pre-shock conditions and shock speed.  In 
\S\ref{sec-results} I show that both fast and intermediate shock 
solutions exist for low shock speeds, and that the topology of the 
phase space prevents the fast shock solution from being found by 
integration from upstream to downstream.  At higher speeds the 
intermediate solutions become unphysical, but the fast solutions 
persist.  I show that marginally coupled grains dominate the grain 
drag, and that their effect on the shock structure is dramatic, and 
examine the effects of an MRN grain-size distribution.  The implications of 
these results are discussed in \S\ref{sec:discussion}, and a summary 
is presented in \S\ref{sec:summary}.

\section{Formulation}
\label{sec-formulation}

The fluid is assumed to be weakly ionized, in the sense that the 
inertia and thermal pressure of the charged components, and the 
collisional coupling between charged species, are unimportant.  This 
is generally an excellent approximation in molecular clouds (Shu 1983; 
Mouschovias 1987), and within C-type shock waves as the fractional 
ionization in dense clouds is less than 1 part in \( 10^6 \).  
Processes that create and destroy different species within the shock 
front, i.e.  ionization, recombination, and chemistry, are important 
(Pilipp et al. 1990), but will be neglected here for simplicity (see \S 
\ref{sec:discussion}).  Further, the pressure in the neutral component 
is set to zero, a justifiable approximation in C-type shocks where the 
dynamics of the flow is weakly dependent on the gas pressure because 
the neutral component is supersonic throughout the shock front (Wardle 
1991b).

Although the fluid temperatures do not directly influence the 
dynamics, they affect the rate coefficients for elastic scattering 
between the charged species and the neutrals (see [\ref{eq-veff,j}]).  
In addition, the electron temperature affects the coupling of the 
grains to the magnetic field because the grain charge is determined by 
the sticking of electrons to grain surfaces (eq.  
[\ref{eq-graincharge}]).  However, I assume that the rate coefficients 
and the grain charge are constant throughout the shock front (with 
values chosen to be representative of the shock interior) so that 
equations for the temperatures need not be integrated.

The equations describing shock structure developed in \S 
\ref{subsec-shock-structure-eqns} are therefore generally simpler than 
those of Pilipp \& Hartquist (1994), who explicitly followed the fluid 
temperatures, grain charging, and chemistry.  They are more general in 
allowing an arbitrary number of charged species, and in not assuming 
that the magnetic flux is frozen into the electron fluid.  The latter 
has little effect for the densities of order \( 10^6 \percc \) 
relevant here, but will allow the equations to be applied at higher 
densities (in excess of \( 10^9 \percc \)) that may be relevant for 
models of shocks associated with maser regions or of accretion shocks 
onto protostellar discs (Neufeld \& Hollenbach 1994) where the 
electron density is low, and ions and PAHs become decoupled from the 
magnetic field.

The adopted properties of the charged species present in the gas -- 
ions, electrons, negatively charged grains and PAHs -- are discussed in 
\S \ref{subsec-charged-species}.  

\subsection{The equations for shock structure}
\label{subsec-shock-structure-eqns}

Each component of the fluid is characterised by particle mass \(m\) and 
charge \(Ze\), mass and number densities \(\rho\) and \(n\), and velocity 
\(\v\).  Unsubscripted quantities refer to the neutral fluid, and 
subscripts \(i\), \(e\) and \(g\) refer to the ions, electrons, and 
negatively charged grains respectively.  The generic subscript `\(j\)' 
will be used to denote any charged species, i.e.  \(j\in\{i,e,g\}\), and 
the equations are trivially generalised to include additional charged 
species by expanding this set.

The shock front is assumed to be steady and plane-parallel, 
propagating at speed \(v_s\) into a uniform medium of neutral density 
\(\rho_0\) permeated by a magnetic field \(\B_0\) which makes an angle 
\(\theta\) to the shock normal.  We adopt a Cartesian coordinate system 
\((x,y,z)\) in the shock frame (see Figure 1) with the \(z\)-axis parallel 
to the direction of shock propagation, and the pre-shock magnetic field 
lying in the \(x\)-\(z\) plane.  The incoming pre-shock gas then has 
velocity \(v_s {\bf \hat z}\). 
\begin{figure}
\centerline{\epsfxsize=7.5cm \epsfbox{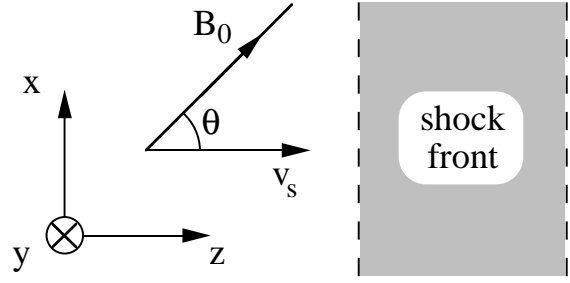}}
\caption{A sketch of the Cartesian reference frame used in this paper.  
In this frame the shock is at rest, and the unshocked material moves 
into the shock along the $z$-axis at speed $v_s$.  The upstream 
magnetic field 
makes an angle $\theta$ to the incoming velocity, and lies in the $ 
x $-$ z $ plane, which is referred to in the text as the `shock 
plane'.}
\end{figure}
 The continuity equations are
\begin{equation}
\rho v_z = \rho_0 v_s
\label{eq-ncontinuity}
\end{equation}
for the neutral fluid, and
\begin{equation}
\rho_j v_{j z} = \rho_{j 0} v_s
\label{eq-jcontinuity}
\end{equation}
for each of the charged species.
The momentum equation for the neutral component of the fluid
is:
\begin{equation}
\ddz{\rho v_z \v } =  \sum_j \gamma_j \rho_j \rho  (\v _j - \v) \,, 
\label{eq-nmom}
\end{equation}
where \(\gamma_j\) is related to the rate coefficient for elastic scattering
between particles of species j and the neutrals, \(<\sigma v>_j\), by
\begin{equation}
\gamma_j = {< \sigma v >_j \over m_j + m}  \,.
\end{equation}
The drift of each charged species through the neutral gas is
determined by a balance between
electromagnetic forces and collisional drag (e.g. Shu 1983):
\begin{equation}
n_j Z_j e \left( \E + {\vj \over c} \cross \B \right) + 
\gamma_j \rho_j \rho  (\v - \vj) = 0 \,.
\label{eq-jforcebalance}
\end{equation}
The inertia of the charged species has been neglected, which
is valid provided that the time scale over which conditions 
change (i.e. the flow time through the shock front)
is long compared to the collisional time with the neutrals.
Equation (\ref{eq-jforcebalance}) can be rewritten as
\begin{eqnarray}
\lefteqn{\v_j = \v  +   c  \left (
\beta_j\frac{(\B\cdot\E')\B}{B^3} \right. + 
\frac{\beta_j^2}{1 + \beta_j^2}{\E'\cross\B\over B^2} + } && \nonumber  \\
&&  \left. \frac{\beta_j}{1 + \beta_j^2}\frac{\B\cross(\E'\cross\B)}{B^3}
\right ) \,,
\label{eq-v_j}
\end{eqnarray}
where 
\begin{equation}
	\E'=\E+\v\cross\B/c
	\label{eq-efluid}
\end{equation}
is the electric field in the frame
comoving with the neutrals, and
\(\beta_j\) is the Hall parameter for species j:
\begin{equation}
\beta_j= {Z_jeB \over m_j c} \, {1 \over \gamma_j \rho}\, ,
\label{eq:hall_j}
\end{equation}
i.e. the product of the gyrofrequency and the time scale for
momentum exchange with the neutral fluid.   Note that the sign of \( 
\beta_j \) is determined by \( Z_j \).
The magnitude of the Hall parameters determines the magnetic flux transport
in a weakly ionized medium (Draine 1980; K\"onigl 1989).  When
\(|\beta_j|\gg 1\), the electric and magnetic stresses individually dominate
the drag force in equation (5) and must almost cancel, thus
\(\E \approx - (\vj \times \B)/c\)
and the particles are tied to the magnetic field lines.  In the opposite
limit, collisions with the neutrals decouple the charged particles from the
magnetic field lines and \(\vj \approx \v\).

Rather than explicitly consider the drift velocity of each charged 
species, it is often more convenient to consider the expression for the 
current density
\begin{equation}
\J = e \sum_j n_j Z_j \vj \
\label{eq-current_density}
\end{equation}
obtained by using (\ref{eq-v_j}) to substitute for \(\v_j\):
\begin{equation}
	\J = \sigma_0 (\Bh \cdot \E')\Bh + 
	\sigma_1 \Bh \cross \E' + 
	\sigma_2 (\Bh \cross \E') \cross \Bh \,,
	\label{eq-jsigmaE}
\end{equation}
where the components of the conductivity tensor are the conductivity 
parallel to the magnetic field,
\begin{equation}
	\sigma_0 = \frac{ec}{B}\sum_{j} n_j Z_j \beta_j \,,
	\label{eq-sigma0}
\end{equation}
the Hall conductivity,
\begin{equation}
	\sigma_1 = \frac{ec}{B}\sum_{j}\frac{n_j Z_j}{1+\beta_j^2}\,,
	\label{eq-sigma1}
\end{equation}
and the Pedersen conductivity
\begin{equation}
	\sigma_2 = \frac{ec}{B}\sum_{j}\frac{n_j Z_j \beta_j}{1+\beta_j^2}
	\label{eq-sigma2}
\end{equation}
(Cowling 1957; Nakano \& Umebayashi 1986).

Now consider Maxwell's equations. 
The induction equation yields \(\curl \E = 0\), implying that \(E_x\) and \(E_y\)
are constant.  Evaluating them far upstream gives:
\begin{equation}
E_x = 0  \,,
\label{eq-ex}
\end{equation}
and
\begin{equation}
E_y =  -{v_s\over c} B_{0x} \,.
\label{eq-ey}
\end{equation}
Finally, \(\div \B = 0\) implies that
\begin{equation}
B_z = B_{0z} \, . 
\label{eq-bz}
\end{equation}
Gauss' law for the
electric field is replaced by the condition for charge neutrality:
\begin{equation}
\sum_j Z_j n_j = 0 \,.
\label{eq-charge-neutrality}
\end{equation}
Ampere's law reduces to
\begin{equation}
{d B_x \over dz} = {4\pi\over c} J_y \, ,
\label{eq-dbxdz}
\end{equation}
\begin{equation}
{d B_y \over dz} = -\,{4\pi\over c} J_x \,, 
\label{eq-dbydz}
\end{equation}
and
\begin{equation}
J_z = 0 \,.
\label{eq-jz=0}
\end{equation}
Equation (\ref{eq-jz=0}) is not independent as eqs 
(\ref{eq-jcontinuity}) and (\ref{eq-charge-neutrality}) guarantee 
\(J_z=0\) upon noting that \(v_{jz}=v_s\) far upstream.

The derivatives in equation (\ref{eq-nmom}) can be eliminated using 
(\ref{eq-jforcebalance}), (\ref{eq-current_density}), (\ref{eq-bz}) 
and (\ref{eq-dbxdz})--(\ref{eq-jz=0}) to obtain expressions for the 
conservation of the \( x \), \( y \), and \( z \) components of the 
total momentum flux:
\begin{equation}
\rho_0 v_s v_x - {B_x B_z \over 4 \pi} = -{B_{0x} B_z \over 4 \pi} \,,
\label{eq-xmomflux}
\end{equation}
\begin{equation}
\rho_0 v_s v_y - {B_y B_z \over 4 \pi} = 0 \,,
\label{eq-ymomflux}
\end{equation}
and
\begin{equation}
\rho_0 v_s v_z + {B_x^2 + B_y^2 \over 8 \pi} = \rho_0 v_s^2 +
{B_{0x}^2 \over 8 \pi}  \,.
\label{eq-zmomflux}
\end{equation}

Thus we have a system of two ordinary differential equations in \(B_x\) 
and \(B_y\), (\ref{eq-dbxdz}) and (\ref{eq-dbydz}), along with the 
algebraic relations (\ref{eq-ncontinuity}), (\ref{eq-jcontinuity}), 
(\ref{eq-v_j}), (\ref{eq-current_density}) 
(\ref{eq-ex})--(\ref{eq-charge-neutrality}), and 
(\ref{eq-xmomflux})--(\ref{eq-zmomflux}).

\subsection{Charged species}
\label{subsec-charged-species}

The pre-shock abundances of charged species are determined by a balance 
between cosmic-ray ionization and recombination occuring in the gas 
phase or on grain surfaces, and are therefore sensitive to the adopted 
grain model (Nishi et al. 1991).

When grains are of a single size \( a  \ga 0.1 \mu \), 
recombination occurs primarily in the gas phase for \( \nh\la 10^{10} 
\percc \), yielding a fractional ionization
\begin{equation} 
{n_i\over \nh } \approx 10^{-8} \left ({\nh \over 10^6 \percc}\right )^{-1/2} 
\end{equation} 
and an almost identical electron abundance (Elmegreen 1979; Nakano \& 
Umebayashi 1986).  The grains in the pre-shock gas carry a single 
negative charge, and assuming the the total mass of grains is 0.01 
that of hydrogen, the grain abundance is \( n_g/\nh= 2.5 \ee -14 \) 
for \( a=0.4 \mu \) (Draine et al. 1983), or \( 1.6 \ee 
-12 \) for \( a=0.1 \mu \).  Under these conditions, the charging of 
the grains within the shock front proceeds by the sticking of 
electrons from the gas phase, and is limited by the Coulomb repulsion 
of incoming electrons by the grain.  The resulting mean 
charge depends on the grain radius and the electron temperature 
(Draine 1980):
\begin{equation}
Z_g \approx {-4 k T_e a \over e^2 } \approx -4.8 \left
({T_e\over200 \u K }\right ) \left ({a\over 10^{-5} \u cm }\right ) 
\,,
\label{eq-graincharge}
\end{equation}
valid for \( |Z_g| \gg 1 \).  The charges of individual grains 
fluctuate about the mean, but at any instant most are within 1 unit 
of the mean (Elmegreen 1979).

However, grains possess a power-law grain-size distribution 
(Mathis, Rumpl \& Nordsieck 1978; Draine \& Lee 1984),
\begin{equation}
\frac{dn}{da} = A\nh a^{-3.5}
\label{eq-mrn}
\end{equation}
where \(n_(a)\) is the number density of grains with radii smaller than 
\(a\), \(A=1.5\ee -25 \ut cm -2.5 \), and the distribution runs from grain 
radii between lower and upper bounds \( a_1= 50 \)\AA\ and \( 
a_2=2500\)\AA.  The lower bound is poorly constrained by observations,
and the upper 
bound is uncertain in dense clouds, where grains are enlarged by
icy mantles.  In addition PAHs with \(a\sim 3 \)\AA\ and \( n/\nh \approx 2\ee -7 \) 
(L\'eger \& Puget 1984; Puget \& L\'eger 1989), may be present, 
either as a distinct population or as the tail of an MRN distribution 
with \( a_1 \approx 3 \AA \).  For the sake of definiteness, I shall 
treat the PAHs as a distinct population.  The MRN size distribution 
and the presence of PAHs increase the grain surface area and reduces 
the fractional ionization in the gas and the electron density 
by several orders of magnitude (Nishi et al. 1991; Kaufman 
\& Neufeld 1996a).  The electron density is comparable with, or less 
than, the density of grains and the charging of grains by electron 
sticking within the shock front is inhibited.  Thus the grain charge 
remains at \( Z_g=-1\), independent of \( T_e \) in this case.

These two cases are considered here. 
First, for comparison with Pilipp \& Hartquist (1994), I consider 
large grains of a single size.  The charged species in this case are 
ions, electrons and negatively charged grains.  The effect of an MRN 
grain-size distribution and PAHs are also considered.  In this case 
the charged species are ions, electrons or PAHs, and negatively 
charged grains.  The grains are still considered to be of a single 
size, but the grain size and abundance are chosen so that the 
contribution to the conductivity of the gas approximates that of the 
entire grain-size distribution (see \S\ref{subsec-grain-size-dist}).

I assume that the ions are singly charged and have mass \(30 m_{\rm 
H}\), as is typical for molecular or metal ions (e.g.  HCO\(^{\rm +}\), 
Mg\(^{\rm +}\)).  Grains are taken to be spherical, all of the same 
radius \( a \), with internal density \(2.5 \u g \ut cm -3 \) and total 
mass 0.01 of the mass in hydrogen.

The rate coefficient for elastic scattering with the neutrals,
\(<\sigma v>_j\), depends in general on the fluid temperatures and the 
drift speed \( |\v_j - \v| \).  The scattering cross-section for ions has 
a \(1/v\) dependence for ions with drift speeds below  20 
\(\kms\), and the 
rate coefficient for ion-neutral scattering is constant,
\(<\sigma v>_i \approx 1.6 \ee -9 \ut cm 3 \ut s -1 \), yielding a 
Hall parameter
\begin{equation}
\beta_i \approx
4600  \left({B \over 1\u mG } \right)
\left( { n_{\rm H} \over 10^{6} \ut cm -3 } \right)^{-1}
\end{equation}
The PAHs have geometric cross-sections \( \pi a^2 \approx 3\ee -16 
(a/\mathrm{\AA})^2 \ut cm -2 \), of order the polarization 
cross-section for elastic scattering from neutrals, \( 1.6\ee -15 
\times 10\kms / |\v_g - \v| \).  The PAHs are therefore treated as 
negatively charged ions, with \( <\sigma 
v>_{\mathrm{PAH}} = <\sigma v>_i \), and \( \beta_{\mathrm{PAH}} = - 
\beta_i \).

The elastic scattering cross-section for grains is the geometric
cross-section, taken to be \(\sigma_{g} = \pi a^{2} \), while
for electrons,  \( \sigma_{e} \approx 1 \ee -15 
\ut cm 2 \) at energies below a few eV (Gilardini 1972).
As \( \sigma_j \) is energy-independent for both these species,
the rate coefficient may be written as \( \sigma_j 
u_{j} \) where the effective velocity is
\begin{equation}
	u_{j} = \left[
\frac{128}{9\pi} \left(\frac{kT}{m} + \frac{kT_j}{m_j}\right) + |\v_j - \v|^2 
	\right]^{1/2}
	\label{eq-veff,j}
\end{equation}
(Draine 1986).
The rate coefficient for electron scattering is determined 
by the electron thermal velocity, which is far in excess of both the 
electron-neutral drift speed and the neutral thermal velocity.  The 
effective relative velocity between electrons and neutrals may 
therefore be written
\begin{equation}
u_{e} =\left(\frac{128kT_{e}}{9\pi m_{e}}\right)^{1/2}\,,
\label{eq-veff,e}
\end{equation}
and the electron Hall parameter is
\begin{equation}
\beta_e \approx
-2.5\ee 6  \left({B \over 1\u mG } \right)
\left( { n_{\rm H} \over 10^{6} \ut cm -3 } \right)^{-1}
\left(\frac{T_{e}}{200 \u K }\right)^{-1/2}
\end{equation}
The grain thermal velocity is negligible compared with that for the 
neutrals, and
\begin{equation}
u_{g}=
\left(\frac{128kT}{9\pi m} + |\v_g - \v|^2 \right)^{1/2} \,.
\label{eq-veff,g} 
\end{equation}
The grain Hall parameter is then [eq. (35) of Draine (1980)]:
\begin{eqnarray}
\lefteqn{\beta_g \approx 0.22\, Z_{g} \left (\frac{B}{\u mG }\right)
\left(\frac{\nh}{10^6 \ut cm -3 }\right )^{-1} 
\left ( \frac{a}{10^{-5}\u cm }\right )^{-2} \times } && \nonumber \\
&& \left(\frac{u_g}{\kms}\right)^{-1}\,.
\label{grainHall1}
\end{eqnarray}

The rate coefficient for grain-neutral scattering depends on the 
neutral gas temperature and the grain-neutral drift speed.  The grain 
charge and the electron scattering coefficients depend on the electron 
temperature.  A calculation of the 
fluid temperatures is beyond the scope of the present work, so we 
assume that \( u_e \), \( u_g \) and \( Z_g \) are constant throughout 
the flow, being fixed at values representative of conditions within 
the shock front.  In particular, for the weak shock models (\( v_s / 
v_A = 1.5 \)) I adopt \( T=100 \u K \), \( T_e=200 \u K \), and \( |\v 
- \v_g| = 1 \kms \); and for the strong shock models (\( v_s / v_A = 
10 \)), \( T=1000 \u K \), \( T_e=2000 \u K \), and \( |\v - \v_g| = 
10 \kms \).  The grain charge is either calculated from 
(\ref{eq-graincharge}) using the adopted values of \( T_e \) or is set 
to \( -1 \), depending on the chosen grain model.

\section{Analysis}
\label{sec-analysis}

The shock structure is determined by the pair of differential 
equations (\ref{eq-dbxdz}) and (\ref{eq-dbydz}) for \(B_x\) and 
\(B_y\) as a function of \(z\), and the associated algebraic 
equations.  There is no explicit reference to \(z\), so 
given a choice of parameters, each shock solution can be represented 
as an integral curve in a phase space with coordinates \(B_x\) 
and \(B_y\), i.e.  as a plot of \(B_y(z)\) vs \(B_x(z)\).  This proves 
to be extremely useful in understanding the relationship between 
different shock solutions.  Trajectories representing shock 
solutions begin and terminate at points representing upstream or 
downstream states, for which
\begin{equation}
\ddz{B_x} = \ddz{B_y} = 0
\end{equation}
(i.e.  stationary points).  The jump conditions, discussed in \S 
\ref{subsec-jump-conditions}, show that there are three stationary 
points, all lying on the \(B_x\) axis, which are denoted by U, F, and I 
in increasing order of \(|B_x|\) (see Fig. \ref{fig-int-phase-space}).  
These points represent the upstream 
state, and possible downstream states corresponding to the fast and 
intermediate shocks respectively.  The topology of the trajectories 
near the stationary points is obtained through a linear analysis in 
\S\ref{subsec-classification}, where it is shown that U, F, and I are 
a source, saddle, and sink respectively.   Thus it is not possible to 
find a fast shock trajectory by starting at the upstream state and 
integrating through the shock front towards the downstream state F; 
the trajectory should be integrated in the reverse direction from F to 
U.

To actually integrate equations (\ref{eq-dbxdz}) and (\ref{eq-dbydz}), 
the \(x\) and \(y\) components of the current density \({\bf J}\) must 
be determined from the algebraic relations given \(B_x\) and \(B_y\) 
at some point in the shock front.  The expressions for total momentum 
flux (\ref{eq-xmomflux})--(\ref{eq-zmomflux}) allow \(\v\) to be 
expressed in terms of \(\B\), the neutral continuity equation then yields 
\(\rho\), and the Hall parameters \(\beta_j\) can then be calculated 
for each charged species.  If \(E'_z\) were known then the charged 
particle drifts, and hence the current density, could be calculated 
from eq.  (\ref{eq-v_j}).  Actually \(E'_z\) is determined from the 
charge neutrality condition (\ref{eq-charge-neutrality}), by substituting 
the \(z\)-component of (\ref{eq-v_j}) and using (\ref{eq-jcontinuity}) 
to relate \(\rho_j\) and \(v_{jz}\) (\S 
\ref{subsec-charge-neutrality}).  The presence of \( N \) charged 
species allows \( N-1 \) solutions for \( E'_z \) that yield drift 
velocities consistent with the charge neutrality condition at each 
point within the shock front.  The choice of solution is dictated by 
requiring that the electric field vanish in the neutral rest frame 
ahead of (or behind) the shock front and be everywhere continuous.

Several regions of the phase space can be excluded on the grounds that 
the density of the neutrals, or of a charged species, becomes negative 
(\S \ref{subsec-forbidden}).  In particular, when \( B^2 \) becomes 
too large, the ram pressure must become negative (see eq 
[\ref{eq-zmomflux}]).  The stationary point I becomes unphysical in this 
manner when the shock velocity is too large to allow intermediate 
shocks (see \S\ref{subsec-jump-conditions}).  When this is the case, 
there is also a locus in phase space where the ion and electron 
densities simultaneously diverge.  However, it will turn out that the 
fast shock solutions \emph{avoid} this locus.

The effect of an MRN grain-size distribution is examined in \S 
\ref{subsec-grain-size-dist}, where it is shown that it can be 
approximated by a single-size grain model with appropriately chosen 
`effective' grain Hall parameters and abundances.

One interesting feature of the shock solutions is that within the shock 
front, the fluid motions and magnetic field are not restricted to lie 
in the plane containing the shock normal and the pre-shock magnetic 
field.  The sense of motion out of the plane (i.e.  `up' or 
`down') is determined by the asymmetry in the elastic scattering 
properties of species of different sign (i.e.\( |\sigma_1|\ne 0 \), 
see eq (\ref{eq-sigma1}) ).  Even if the Hall parameters of all 
charged species satisfy \( |\beta_j| \la 1 \), fast shocks remain 
coplanar if the properties of negatively and positively charged 
species are identical apart from the difference in sign.  A criterion 
for this effect to be important is developed in 
\S\ref{subsec-criteria}.  In essence, there should be `sufficient' 
poorly-coupled charged particles, with the charge on the particles 
being preferentially of one sign.

Finally, in \S\ref{subsec-criteria} I discuss the dimensionless 
parameters that determine the shock structure, and outline how the 
calculations presented in \S \ref{sec-results} were carried out.

\subsection{Jump conditions}
\label{subsec-jump-conditions}

The boundary conditions for equations (\ref{eq-dbxdz}) and 
(\ref{eq-dbydz}) are that the derivatives of \( B_x \) and \( B_y \) 
should vanish far away from the shock front.  This implies that the 
current density, and hence (via eq [\ref{eq-jsigmaE}]) the electric 
field in the rest frame of the neutrals also vanish at \( z=\pm \infty 
\).  In turn, (\ref{eq-v_j}) then guarantees that \(\v_j = \v\) far 
upstream and downstream of the shock front.  This is, of course, 
expected, and shows that the MHD jump conditions which relate the 
upstream and downstream states can be obtained by regarding the 
weakly ionized fluid as a plasma of density \(\rho\) and velocity \(\v\), 
and equating the fluxes of mass and momentum of the combined fluid on 
either side of the front.  The ionized component does not contribute 
significantly to the fluid density or mass flux, but does contribute 
to the momentum flux through the magnetic field (see eqs 
[\ref{eq-xmomflux}]--[\ref{eq-zmomflux}]).  An energy condition is not 
required, as it is assumed that \( T=0 \).

In general, the MHD jump conditions imply that the pre-shock and 
post-shock magnetic fields and velocities are coplanar, and three broad 
classes of shocks can be distinguished on the basis of the behaviour 
of the field component transverse to the shock velocity, which either 
increases, decreases, or changes sign in going from upstream to 
downstream in fast, slow, and intermediate shocks respectively (e.g.  
Cowling 1976; Kennel et al. 1989).

The signal speeds in the fluid play a key role in determining the
different kinds of shock the medium can support.  The combined fluid
supports fast, intermediate and slow waves with signal speeds for
propagation in the direction normal to the shock front of
\begin{equation}
v_{\mathrm{fast}} = \left( {B^2 \over 4 \pi\rho} \right)^{1/2} = v_A
\end{equation}
where \(v_A\) is the Alfv\'en speed,
\begin{equation}
v_{\mathrm{int}} = \left ( {B_z^2 \over 4 \pi\rho} \right)^{1/2} = v_A \cos \theta
\,,
\end{equation}
and
\begin{equation}
v_{\mathrm{slow}} = 0 \,,
\end{equation}
respectively.  The assumption that 
the plasma is cold eliminates shock transitions that have downstream 
velocities below the slow speed (i.e.  slow shocks and three classes 
of intermediate shock, see Kennel et al. 1989).

Far away from the shock front,
\begin{equation}
\E = {-\v / c} \cross \B
\end{equation}
which implies that
\begin{equation}
v_y B_z - v_z B_y = 0\,,
\label{eq-vybz}
\end{equation}
and
\begin{equation}
v_z B_x - v_x B_z = v_s B_{0x} \,.
\label{eq-vzbx}
\end{equation}
Equations (\ref{eq-ncontinuity}) and 
(\ref{eq-xmomflux})--(\ref{eq-zmomflux}) for mass and momentum flux conservation
also hold and combined with (\ref{eq-vybz}) and (\ref{eq-vzbx}) yield a
complete set of jump conditions.

Equations (\ref{eq-ymomflux}) and (\ref{eq-vybz}) both express \(v_y\) 
downstream in terms of \(B_y\), implying that either \(B_y = 0\), or that
\begin{equation}
v_z = {B_z^2 \over 4 \pi \rho_0 v_s} \,.
\end{equation}
In the latter case, the \(z\)-components of the pre-shock and post-shock
velocities are equal to the intermediate speed, and the transverse
components of the magnetic field and fluid velocity change direction
in such a way as to conserve the transverse momentum flux.  There is
no compression or entropy increase across such a transition, which
exists in ideal MHD as a `rotational discontinuity' (e.g. Cowling
1976).  In the non-ideal case it is unsteady, broadening diffusively
with time (Wu 1988b), and we shall not discuss it further here.

Thus we set \(B_y = 0\) and \(v_y = 0\) and use (\ref{eq-xmomflux}) 
and (\ref{eq-zmomflux}) to eliminate \(v_x\) and \(v_z\) in favour of 
\(B_x\) in (\ref{eq-vzbx}), to obtain a cubic in \(B_x\) whose 
solutions satisfy the jump conditions.  The upstream state is a 
solution, so \((B_x-B_{0x})\) factors out of the cubic to yield a 
quadratic
\begin{equation}
 B_x^2 + B_{0x}B_x +2\left(B_z^2 - 4\pi\rho_0 v_s^2\right) = 0 \,
\end{equation}
with positive and negative roots that are candidate downstream states for 
fast and intermediate shock transitions respectively.

The roots correspond to physically acceptable downstream states if 
there is compression across the shock front, i.e.  for \(0 < v_z < 
v_s\).  From eq (\ref{eq-zmomflux}), the condition on \(B_x\) is:
\begin{equation}
1 < \left({B_x\over B_{0x}}\right)^2 < 1+{8 \pi\rho_0v_s^2 \over B_{0x}^2} \,.
\end{equation}
When \(v_s /v_A > 1\), the positive root yields a fast shock, and when 
\(\cos \theta < v_s/v_A < {\sqrt 2}\, \cot \theta\) the negative root 
yields an intermediate shock (c.f.  Kennel et al. 1989).  
The regime in which intermediate shocks are allowed is sketched in 
Fig.  2.
\begin{figure}
\centerline{\epsfxsize=10.0cm \epsfbox{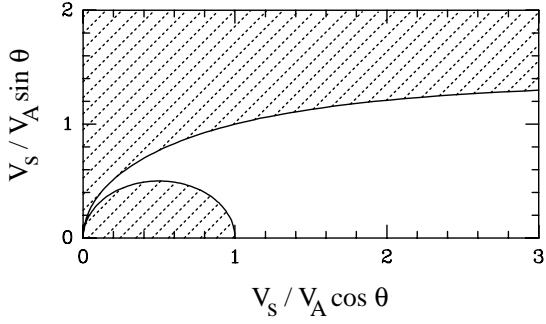}}\vskip -5.5cm
\caption{A polar plot of the conditions for the
existence of intermediate shocks.  A point on the plane represents a
shock velocity, with polar radius representing the shock speed
$v_s$, and polar angle representing the angle $\theta$ between the 
pre-shock magnetic field and the shock normal.  Intermediate shocks 
do not exist in the shaded region.}
\end{figure}

\subsection{Classification of the stationary points}
\label{subsec-classification}

The stationary points can be classified as source, sink or saddle by 
linearizing the equations for shock structure around each point.  I 
use eqs.  (1) and (16)--(18) for the conservation of neutral mass flux 
and total momentum flux, but rather than explicitly considering the 
drift velocities of the charged species given by (\ref{eq-v_j}), it is 
more convenient to use (\ref{eq-jsigmaE}).  Since \(\E'=0\) outside 
the shock front, the linearised version of this equation is
\begin{equation}
	\delta\J = \sigma_0 (\Bh \cdot \delta\E')\Bh + 
	\sigma_1 \Bh \cross \delta\E' + 
	\sigma_2 ( \Bh \cross \delta\E') \cross \Bh
	\label{eq-dJ}
\end{equation}
Assuming that the perturbations are steady, independent of \( x \) and 
\( y \), and vary exponentially in \(z\) as \(e^{\lambda z}\), the 
resulting dispersion relation is
\[
\left [ (v_s^2 - v_A^2\cos^2\theta) \mu - 
\frac{(\sigma_1^2+\sigma_2^2)\sin^2\theta + \sigma_0\sigma_2\cos^2\theta}
{\sigma_0} \right ] \times
\]
\begin{equation}
\qquad \left [ (v_s^2-v_A^2) \mu - \sigma_2 \right ] + \sigma_1^2\cos^2\theta = 0 \,,
\label{eq-dispersion_relation}
\end{equation}
where
\begin{equation}
\mu = \frac{4 \pi v_A (\sigma_1^2+\sigma_2^2)}{c^2 v_s^2 \lambda} \,.
\end{equation}
  The nature of the 
stationary point is determined by the sign of the real part of 
\(\lambda\), which is the same as that of \(\mu\).  Noting that 
\(\sigma_0\), \(\sigma_2 > 0\), and using \(v_s > v_A\), we find that 
the real parts of the roots of the dispersion relation are positive, 
so the upstream state is a source.

The dispersion relation can be applied to the downstream states F and 
I by noting that a boost from the shock frame to a frame in which the 
shock is stationary and the \emph{downstream} fluid velocity is normal 
to the shock does not affect the assumptions of steady state and the 
\(e^{\lambda z}\) dependence of the perturbations.  Thus 
(\ref{eq-dispersion_relation}) can be applied downstream on 
replacing \( v_s \) by \( v_z \), \( v_A \) by the downstream Alfv\'en speed, 
and \( \theta \) by the value of \( \cos^{-1}(B_x/B) \) far downstream.  
Re-evaluating the dispersion relation at F (for which \(v_A > v_z > 
v_A\,{\rm cos}\,\theta\)) shows that the roots there are real and of 
opposite signs, and at I (where \(v_z < v_A\,{\rm cos}\,\theta\)) the 
roots have negative real part.  Thus F is a saddle point, and I is a 
sink.

Finally, we note that at U and I, when
\begin{eqnarray}
&& (v_z^2-v_A^2)\frac{\sigma_1^2+\sigma_2^2}{\sigma_0^2}
    - v_z^2 \frac{\sigma_2}{\sigma_0}\nonumber \\
&&+ 2 (v_z^2-v_A^2)^{1/2}(v_z^2 - v_A^2\cos^2 \theta)^{1/2}
\frac{\cos\theta}{\sin^2\theta}\frac{|\sigma_1|}{\sigma_0}  > 0 \,
\label{eq-spiralcriterion}
\end{eqnarray}
the roots of the dispersion relation, eq. (\ref{eq-dispersion_relation})
are complex conjugates and trajectories spiral out of U and into I.  
Typically, \( \sigma_0^2 \gg \sigma_1^2 + \sigma_2^2 \) and \( v^2 \gg 
v_A^2 \), and the criterion becomes
\begin{equation}
	|\sigma_1| > \frac{\cos\theta}{\sin^2\theta} \sigma_2 \,,
	\label{eq-spiralcriterion2}
\end{equation}
or \( |\beta_g| \la 0.3\) for \( \theta=30\degr \).

\subsection{The charge neutrality condition}
\label{subsec-charge-neutrality}

To actually integrate equations (\ref{eq-dbxdz}) and (\ref{eq-dbydz}), 
the \(x\) and \(y\) components of the current density \({\bf J}\) must 
be determined given \(B_x\) and \(B_y\).  This is achieved as follows.  
The expressions for total momentum flux 
(\ref{eq-xmomflux})--(\ref{eq-zmomflux}) allow \(\v\) to be expressed 
in terms of \(\B\), the continuity equation then yields \(\rho\), and 
thus the Hall parameters \(\beta_j\) can be calculated for each 
charged species.  At this point, \( E'_x \) and \( E'_y \) can be 
found from (\ref{eq-efluid}), (\ref{eq-ex}) and (\ref{eq-ey}), so that 
if \(E'_z\) (or equivalently \( E_z \)) were known then the charged 
particle drifts, and hence the current density, could be calculated 
from eq.  (\ref{eq-v_j}).  Actually \(E'_z\) is determined by the 
charge neutrality condition (\ref{eq-charge-neutrality}), by 
substituting the \(z\)-component of (\ref{eq-v_j}) and using 
(\ref{eq-jcontinuity}) to relate \(\rho_j\) and \(v_{jz}\):
\begin{equation}
0=\sum_j \frac{n_{j0} Z_j}{v_{jz}} \,.
\label{eq-charge-neutralityb}
\end{equation}
Noting that eq. (\ref{eq-v_j}) implies that \(v_{jz}\) can be written in the form
\begin{equation}
v_{jz} = n_{j0} Z_j (p_j E'_z + q_j)
\end{equation}
where
\begin{equation}
p_j = \frac{1}{n_{j0}Z_j} \frac{\beta_j(1+\beta_j^2 B_z^2 / B^2)}{(1+\beta_j^2)}
\end{equation}
and
\begin{eqnarray}
\lefteqn{
q_j = \frac{1}{n_{j0}Z_jv_s} \left [ v_z +
\frac{c\beta_j^2}{1+\beta_j^2}\frac{(\E'\cross\B)_z}{B^2} + \right. } &&\nonumber \\
 &\qquad&\qquad \qquad\left. \frac{c\beta_j^3}{1+\beta_j^2}
 \frac{(\B\bmath{\cdot}\E')B_z}{B^3} \right ] \,,
\end{eqnarray}
eq. (\ref{eq-charge-neutralityb}) can be recast as
\begin{equation}
\sum_j \frac{1}{p_j E'_z + q_j } = 0 \,.
\label{eq-charge-neutralityc}
\end{equation}
In this expression, \( p_j \) and \( q_j \) are known quantities, and 
the task is to determine \( E'_z \).  If N is the number of charged 
species, the sum regarded as a function of \(E'_z\) has N poles at 
\(E'_z = E_j\) where \(E_j = -q_j/p_j\).  If the poles are ordered so 
that \(E_1 < E_2 \ldots < E_n\), then there is a zero between each 
pair \(E_j\), \(E_{j-1}\).  Thus locally there are several values of 
\(E'_z\) which will produce particle drifts with no net current.  The 
correct choice is dictated by the requirement that 
\(E'_z(z)\) be smooth and go to zero ahead of and behind the shock.  

Multiplying (\ref{eq-charge-neutralityc}) by \( \prod_{j} (p_j E'_z + 
q_j) \) yields a polynomial in \(E'_z\) which for N charged species 
has N-1 solutions.
For the case \( N=3 \) relevant to this paper, eq.
(\ref{eq-charge-neutralityc})
yields a quadratic in \(E'_z\):
\begin{equation}
a_2 {E'_z}^2 + a_1 E'_z + a_0 = 0 \,,
\end{equation}
where the required root is
\begin{equation}
	E'_z=\frac{-a_1-\sqrt{a_1^2-4 a_0 a_2}}{2 a_2} \,.
\label{eq-Ez}
\end{equation}

\subsection{Forbidden regions of phase space}
\label{subsec-forbidden}
Several regions in the \( B_x \)--\( B_y \) phase space 
are forbidden in the sense that the density of one or more species 
formally becomes negative. 
 
Firstly, the magnetic field strength is limited by :
\begin{equation}
\frac{B_x^2 + B_y^2}{B_0^2}  <  \frac{2v_s^2}{v_A^2} + \sin^2 \theta \,
		\label{eq-neutral-locus}
\end{equation}
otherwise eq (\ref{eq-zmomflux}) implies that
\(v_z\) would need to be negative to conserve the total z-momentum 
flux.   As the magnetic field approaches the limiting value, \( v_z 
\) tends to zero and the neutral density diverges.

An additional restriction is imposed by requiring that \( v_{jz} > 0 \) 
for each charged species.  The locus \( v_{jz}=0 \) corresponds to 
diverging particle density (see eq. [\ref{eq-jcontinuity}]), and the charge neutrality requirement 
forces the density of an oppositely charged species 
to also diverge there.  Thus for the system of ions, electrons, and 
negatively charged grains considered here, there are two loci:
\( v_{iz}=v_{ez}=0 \), and \( v_{iz}=v_{gz}=0 \).  Which of 
these loci is relevant is intimately related to the choice of \( 
E'_z \) as determined by the charge neutrality condition.  As the 
ions and electrons are better tied to the field lines than the grains,
the choice of \( E'_{z} \) shows that it is the first 
condition that is relevant.
For \(|\beta_{e}|\) and \(\beta_{i} \gg 1\), the locus
\( v_{iz}=v_{ez}=0  \) is given by
\[
\left(\frac{B_x}{B_0}+\frac{v_s^2-v_{\rmn{int}}^2}{v_{\rmn{int}}^2}\sin\theta\right)^2
+ \left(\frac{B_y}{B_0}\right)^2 =
\]
\begin{equation}
	= \frac{v_s^4}{v_{\rmn{int}}^4}\left(\sin^2\theta - 2 
	\frac{v_{\rmn{int}}^2}{v_s^2}\right) \,,
	\label{eq-locus}
\end{equation}
a circle centred on a point on the negative \(B_x\) axis.  The 
region enclosed \emph{within} the locus is unphysical, as the 
electron and ion densities become negative there.  This locus exists 
only if the right-hand side of eq. (\ref{eq-locus}) is positive, i.e.
when \( v_{s}>\sqrt{2} v_{A} \rmn{cot}\, \theta \).  This is 
precisely the criterion for the non-existence of the intermediate 
shocks, so the unphysical region in the \(B_x\)--\(B_y\) plane is only 
relevant for fast shock waves.

The constraint that \( v_{jz}>0 \) arises from the continuity 
equations for the charged species and the requirement that the charged 
particle densities should remain positive.  To some degree this is an 
artifact of the neglect of ionization and recombination processes 
ocurring within the shock front, recombinations serving to limit the 
divergence of the charged particle densities.  In principle, charged 
particles could drift backwards at some point within the shock, with a 
steady state being maintained by recombinations around the points 
where \( v_{jz}=0 \).  In any case, the increase in charged particle 
density increases the currents that can flow within the shock front 
and prevent the trajectories from crossing into the unphysical zone 
(see \S 4.2).

\subsection{The grain-size distribution}
\label{subsec-grain-size-dist}

The majority of studies (e.g.  Draine et al. 
1983; Pilipp \& Hartquist 1994) assumed grain sizes of 0.4 \(\mu\) and 
0.1 \( \mu \) respectively.  However, the smaller grains in an MRN 
distribution dominate the conductivity of the gas (Nakano \& 
Umebayashi 1986) and determine the shock structure (Kaufman \& 
Neufeld 1996a).  If all grains have the same drift velocity, the 
cross-section for collisions with neutrals scales with grain radius as 
\( a^{2} \), and the differential contribution to the drag by grains 
with radii in the range \( a \) to \( a+da \) scales as \( a^{-1.5} da 
\propto da^{-0.5}\).  (The decrease at larger radii will be a little 
steeper as grains become more strongly tied to the neutral gas.)  The 
dust contribution to the drag force on the neutral gas is therefore 
dominated by the smallest grains, but a broad range of sizes 
contributes significantly to the total drag force.

In principle, the equations of \S\ref{subsec-shock-structure-eqns} can 
be applied in this case with each size grain treated as a different 
species, and the sums over \( j \) converted to integrals over grain 
size.  Such a treatment is complicated by the dependence of \( |\v_g - 
\v| \) on grain size, and lies beyond the scope of the present paper.  
Instead, I develop an approximate treatment in which an effective 
grain size and number density are determined.

The rationale for selecting the `effective' pre-shock grain no.  
density \( n_g \) and Hall parameter \( \beta_g \) is that the grain 
contribution to the components of the conductivity tensor 
perpendicular to \( \B \), i.e. \( 
\sigma_1 \) and \( \sigma_2 \), should match that for the entire size 
distribution.  It is not possible to simultaneously match the 
contribution to the
conductivity parallel to the magnetic field,\( \sigma_0 \), but this 
is dominated in any case by the species that have large Hall 
parameters, i.e.  ions, and electrons or PAHs (see 
[\ref{eq-sigma0}]).  Thus \( n_g \) and \( \beta_g \) must satisfy
\begin{equation}
	\frac{n_g}{1+\beta_g^2} = \int_{a_1}^{a_2} 
	\frac{dn_g/da}{1+\beta_g^2}\,da = N I_1
	\label{eq-sigma1-constraint}
\end{equation}
and
\begin{equation}
	\frac{n_g|\beta_g|}{1+\beta_g^2} = \int_{a_1}^{a_2} 
	\frac{|\beta_g| \, dn_g/da}{1+\beta_g^2}\,da = N I_2
	\label{eq-sigma2-constraint}
\end{equation}
where
\begin{equation}
N={\textstyle \frac{2}{5}} A\nh a_{1}^{-2.5} 
	\approx 3.4 \ee -10 \nh \left(\frac{a_{1}}{50 \mathrm{\AA}}\right)^{-2.5}
	\label{eq-ngrains}
\end{equation}
is the total number of grains with radii between \(a_1\) and 
\(a_2\) apart from a negligible correction factor of \( [
1-(a_1/a_2)^{2.5} ] \),
\begin{equation}
	I_1 = {\textstyle \frac{5}{4}}\beta_1^{-5/4} 
	\int_{\beta_2}^{\beta_1}\frac{\beta^{1/4}}{1+\beta^2} \, d\beta\,,
	\label{eq-I1}
\end{equation}
and
\begin{equation}
	I_2 = {\textstyle \frac{5}{4}}\beta_1^{-5/4} 
	\int_{\beta_2}^{\beta_1}\frac{\beta^{5/4}}{1+\beta^2} \, d\beta\,,
	\label{eq-I2}
\end{equation}
where \( \beta_1 \) and \( \beta_2 \) are the absolute values of the
Hall parameters for 
grains of sizes \( a_1 \) and \( a_2 \) respectively. The integrals 
over \( a \) have been converted to integrals over \( \beta \propto 
a^{-2}\).
Thus 
\begin{equation}
	\beta_g = - I_2 / I_1
	\label{eq-eff-betag}
\end{equation}
and
\begin{equation}
	n_g = (1+\beta_g^2) I_1 N \,.
	\label{eq-eff-ng}
\end{equation}

The integrals for \( I_1 \) and \( I_2 \) are solved analytically 
below, but it is instructive to derive approximate values before 
presenting the exact results.   These can be obtained by setting 
\( 1+\beta^2 \) in (\ref{eq-I1}) and (\ref{eq-I2}) to 1 for \( 
\beta<1 \) and to \( \beta^2 \) for \( \beta > 1 \).  This yields
\begin{equation}
	I_1 \approx \left\{ 
	\begin{array}{ll}
		1 & \beta_2\ll \beta_1 \ll 1
		\\[3pt]
		{\textstyle \frac{4}{3}} \beta_1^{-5/4} & \beta_2 \ll 1 \ll\beta_1  
		\\[3pt]
		{\textstyle \frac{5}{3}} (a_2/a_1)^{3/2}\beta_1^{-2} & 1 \ll \beta_2 \ll \beta_1
	\end{array} \right.
	\label{eq-I1-approx}
\end{equation}
and
\begin{equation}
	I_2 \approx \left\{ 
	\begin{array}{ll}
	{\textstyle \frac{5}{9}} \beta_1 & \beta_2\ll \beta_1 \ll 1  \\[3pt]
		5 \beta_1^{-1} & 1 \ll\beta_1    \\[3pt]
	\end{array} \right.
	\label{eq-I2-approx}
\end{equation}
The approximation is poor for \( I_1 \) when \( \beta_2 \ll 1 
\ll\beta_1 \), as there is a significant contribution for \( \beta 
\) near 1.  Thus the expression given above includes a correction 
factor of \( 1/2 \).  These expressions yield estimates of the 
effective grain parameters:
\begin{equation}
	\frac{|\beta_g|}{\beta_1} \approx \left\{ 
	\begin{array}{ll}
		{\textstyle \frac{5}{9}} & \beta_2\ll \beta_1 \ll 1  \\[3pt]
		{\textstyle \frac{3}{2}} \beta_1^{-3/4} & \beta_2 \ll 1 
		\ll\beta_1  \\[3pt]
		3 (a_2/a_1)^{-3/2} & 1 \ll \beta_2 \ll \beta_1 \,,
	\end{array} \right. 
	\label{eq-betag-approx}
\end{equation}
and
\begin{equation}
	\frac{n_g}{N} \approx \left\{ 
	\begin{array}{ll}
		1 & \beta_2\ll \beta_1 \ll 1  \\[3pt]
		3 \beta_1^{-3/2} & \beta_2 \ll 1 \ll\beta_1  \\[3pt]
		15 (a_2/a_1)^{-3/2} & 1 \ll \beta_2 \ll \beta_1 \,.
	\end{array} \right.
	\label{eq-ng-approx}
\end{equation}

These approximations show that when all of the grains are decoupled 
from the magnetic field (\( \beta_1 \ll 1 \)), the effective Hall 
parameter is roughly half that of the most strongly coupled grains, 
and all of the grains contribute to the noncoplanarity.  When all of 
the grains are coupled (\( \beta_2\gg 1 \)), \( \beta_g \approx 3 
(a_2/a_1)^{1/2} \beta_2\), and the number of participating grains is 
of order one per cent.  However, the most relevant case is the 
intermediate regime, in which the smallest grains are coupled to the 
field, whereas the largest grains are not (\( \beta_2 \ll 1 \ll 
\beta_1 \)), as from (\ref{grainHall1}),
\begin{equation}
	\beta_1 \approx 8.6
	\left (\frac{B/\u mG } {\nh/10^6 \ut cm -3 }\right )
	\left ( a \over{50 \AA }\right )^{-2} \nonumber \\
	\left(\frac{v_{\mathrm{eff}}}{10 \kms}\right)^{-1}\,,
	\label{eq-beta1}
\end{equation}
and \( \beta_2 \approx 2500 \beta_1 \).  Then the effective 
Hall parameter is of order unity, and a tenth of the grains or more 
can be regarded as participating.

\( I_1 \) and \( I_2 \) can be found analytically by making the 
substitution \( x=\beta^{1/4} \) and using the identity
\begin{eqnarray}
	\lefteqn{I_{m,n}(x) \equiv \int_{}^{}\frac{x^{m-1}}{1+x^{2n}} \, dx = }
	\nonumber \\
	& & 	\sum_{k=1}^{n}\left[ \sin m\theta_k 
	\tan^{-1}\left( \frac{x-\cos\theta_k}{\sin\theta_k} \right) - \right.
    \nonumber \\
	& & \left. \frac{1}{2} \cos m\theta_k \ln (1-2x\cos\theta_k+x^2) \right] \,,
    	\label{eq-Imn}
\end{eqnarray}
where
\begin{equation}
	\theta_k = \frac{2k-1}{2n}\pi \,,
	\label{eq-thetak}
\end{equation}
which is valid for \( m \) and \( n \) natural numbers and \( m < 2n 
\) (Gradshteyn \& Ryzhik 1965, 2.146).  Then
\begin{equation}
	I_1 = 5\beta_1^{-5/4} 
	\left[ I_{5,4}(\beta_1^{1/4}) - I_{5,4}(\beta_2^{1/4})\right]
	\label{eq-I1-analytic}
\end{equation}
\begin{equation}
	I_2 = 5\beta_1^{-5/4} 
	\left[ I_{1,4}(\beta_2^{1/4}) - I_{1,4}(\beta_1^{1/4}) +
	\beta_1^{1/4} - \beta_2^{1/4}\right]
	\label{eq-I2-analytic}
\end{equation}

\begin{figure}
\centerline{\epsfxsize=8cm \epsfbox{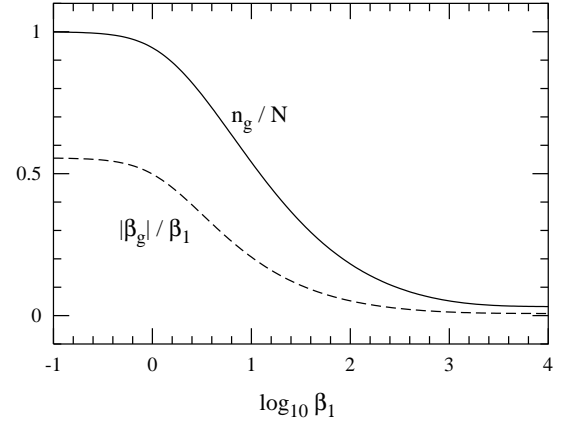}}\vskip 0cm
\caption{The effective single-size grain parameters corresponding to 
an MRN distribution between 50 and 2500 \AA, plotted as a function of 
the magnitude of the Hall parameter of the 50 \AA\ grains, $ \beta_1 $.  
The solid 
curve shows the effective grain no.  density relative to the total 
grain no. density, and the dashed curve shows the effective Hall 
parameter $ \beta_g $ relative to $ \beta_1 $.  $ \beta_1 $ is 
typically in the range 0.1 to 1 in dense clouds.  }
\label{fig-effgrainpars}
\end{figure}

The corresponding values of \( n_g \) and \( \beta_g \) are plotted in 
Figure \ref{fig-effgrainpars}.  As \( \beta_1 \) ranges from 1 to 10, 
\( n_g / N \) decreases from about 1 to 0.5, and \( |\beta_g| \) 
increases from 0.6 to 3.  Thus the effective Hall parameter typically 
lies between 0.1 and 2, and the effective number density is between 
0.5 and 1 of the actual charged grain number density.

\subsection{Criteria for non-coplanarity of fast shocks}
\label{subsec-criteria}

The general criterion is simply \( |\sigma_1| \ga \sigma_2 \).
A rough criterion, assuming that ions and 
electrons are well-coupled is that the contribution of poorly-coupled 
grains to the current is a significant fraction of the current that 
would flow within the shock in the absence of grains.  The former can 
be estimated by assuming that the grain-ion drift is \( 1/2 \) of the 
shock speed, the latter from the magnetic field gradient in a 
perpendicular C-type shock, yielding:
\begin{equation}
\half n_g e |Z_g| v_s \ga
\frac{c}{4\pi} \frac{\sqrt{2} v_s / v_A B_0}{L_{\mathrm{shock}}}
	\label{eq-noncoplanar-current}
\end{equation}
where
\begin{equation}
	L_{\mathrm{shock}}=\frac{\sqrt{2} v_A}{n_i <\sigma v>_{i}}
	\label{lshock}
\end{equation} is the characteristic shock thickness in the absence 
of grains (Wardle 1990).
This criterion can be rearranged to yield:
\begin{equation}
	\frac{n_g|Z_g|}{n_i} \beta_{i} \ga 1 \,.
	\label{eq-noncoplanar-criterion}
\end{equation}

A more rigorous derivation begins by considering the grain 
contribution to the net drag force on the neutrals, or equivalently, 
the Lorentz force per unit volume:
\begin{equation}
	\frac{\J\cross\B}{c} = \frac{\sigma_1 (\B\cross\E')\cross\Bh +
	\sigma_2 \B\cross\E'}{c} \,.
	\label{eq-jcrossb}
\end{equation}
Assuming that \(n_{g}\ll n_{e}\approx n_{i}\), and that \(|\beta_{e}|\gg 
\beta_{i}\gg 1\), then the electron contribution is dominated by the 
ions, and the ion contribution to eq. 
(\ref{eq-jcrossb}) is
\begin{equation}
	\left(\frac{\J\cross\B}{ec} \right)_{\mathrm{i}} \approx
	\frac{n_{i}Z_{i}}{\beta_{i}} \,
	{\B\cross\E'} \,.
	\label{eq-jcrossbi}
\end{equation}
The grains contribute a force
\begin{equation}
	\left(\frac{\J\cross\B}{ec} \right)_{\mathrm{g}} \approx
	\frac{n_{g}|Z_{g}|}{1+\beta_{g}^{2}} \,
	(\Bh\cross\E')\cross\B +
	\frac{n_{g}Z_{g}\beta_g}{1+\beta_{g}^{2}} \, 
	\B\cross\E'
	\label{eq-jcrossbg}
\end{equation}
which dominates the ion drag provided that 
\begin{equation}
	\frac{|n_{g}Z_{g}|}{\sqrt{1+\beta_{g}^{2}}}\ga
	\frac{n_{i}Z_{i}}{\beta_{i}} \,.
	\label{eq-gdragcriterion}
\end{equation}
The grain contribution is non-coplanar when \(\beta_{g}\la 1\), thus 
combining this requirement with 
(\ref{eq-gdragcriterion}) yields (\ref{eq-noncoplanar-criterion}).

The criterion for non-coplanarity, eq.  (75), is the opposite to that 
given by Pilipp \& Hartquist (1994, \S 3).  Their sense of the criterion 
implies significant rotation in the limit of no grains, \( n_g 
\rightarrow 0 \), yet fast shock waves in this limit are planar.  
This was based upon a linear analysis of the shock precursor, and 
derived for the growing mode that is relevant to intermediate shock 
precursors, rather than the fast shocks which the criterion of eq.  
(75) addresses.  Even when the Hall current vanishes -- in the pure 
ambipolar diffusion limit -- rotation in intermediate shock precursors 
is enforced by the requirement imposed by the jump conditions that the 
sign of \( B_x \) changes across the shock front.  Fast shocks, 
however, will not exhibit rotation unless the Hall current is 
important.

\subsection{Numerical solution}
\label{subsec-numerical}
The equations are put in dimensionless
form by using \(v_s\), \(B_0\), \(v_s B_0 /c\) as units of velocity, magnetic
field strength and electric field respectively.
The parameters that appear in the equations are 
\begin{enumerate}
\item \(A=v_s/v_{A}\), the
Alfv\'en number, which determines the strength of the shock,
\item \(\theta\), the angle between the upstream
magnetic field and the shock velocity,
\item \(\beta_{j0}\), the Hall parameters in the
pre-shock gas (eq. [\ref{eq:hall_j}]), which determine whether the field is frozen
into each species (\(\beta_j \ga 1\)), and
\item \( \gamma_j \rho_{j0} / \sum_j \gamma_j \rho_{j0}\), which is a
relative measure of the contribution of species \(j\) to the net drag force 
on the neutral component given equal drift velocities of the charged 
species.
\end{enumerate}
The overall normalisation of the drag forces 
appears in the expression for a characteristic length scale,
\begin{equation}
	L_s = {v_A / \sum_j \gamma_j \rho_{j0} }\,,
	\label{eq-lshock}
\end{equation}
which appears as the unit of length in the dimensionless equations for 
\( dB_x/dz \) and \( dB_y/dz \).  Although this is approximately equal 
to the shock thickness \( L_{\mathrm{shock}} \) when the charged 
species are well coupled (Wardle 1990, see also Smith 1992), it 
overestimates the thickness in the present case, when the grains 
do not drift with the magnetic field.  

Numerical integration of the differential equations forwards 
(backwards) in \(z\) from the upstream (downstream) state yield the 
shock structure, i.e \(B_x\) and \(B_y\) (and hence velocity 
components and fluid densities), as a function of \(z\).  The 
derivatives of \(B_x\) and \(B_y\) vanish far upstream and downstream 
(in the limit that \(|z| \rightarrow \infty\)), so the field 
components have to be perturbed before commencing the integration.  
Because the Hall parameters are extremely large for the ions and 
electrons (\(\approx 10^4\), and \(10^6\) respectively) the equations 
are stiff (this could be removed by assuming that the ions and 
electrons are well-tied to the field lines), so I use the
integration method of Gear (1971).

\section{Results}
\label{sec-results}

In this section, I present solutions for C-type shock waves 
propagating into dense molecular gas.  Initially, a single-size grain 
model is adopted, with grain radius 0.4 \( \mu \) and abundance \( 
n_g/\nh = 2.5\ee -14 \).  The pre-shock density is set to \( \nh = 
10^6 \percc
\), \(B_0 = 1\) mG, and \( \theta = 30 \degr \), and the pre-shock ion 
density is \(n_i/\nh = 10^{-8}\).

Mildly super-Alfv\'enic shocks with \( v_s/v_A = 1.5 \) are considered 
in \S\ref{subsec-weak-shocks}.  There is a family of intermediate 
shocks and a single fast shock solution at this shock speed.  Stronger 
shocks, with (\( v_s/v_A = 10 \)), for which only the fast shock 
solution exists, are considered in \S\ref{subsec-strong-shocks}.  For 
the chosen parameters, the fast shock is significantly non-coplanar.  
The effect of adopting a grain size of 0.1 \( \mu \), with a larger 
grain abundance, and of increasing the pre-shock density to \( 10^7 
\percc \) are also examined.

Finally, the dynamical effects of an MRN grain-size distribution are 
examined in \S\ref{subsec-grain-size-results}, where solutions are 
presented for cases with and without PAHs present.

\subsection{Weak fast and intermediate shocks}
\label{subsec-weak-shocks}

For these solutions we adopt 100 and 200 K as the effective values of 
neutral and electron temperatures respectively, consistent with the 
results of Pilipp \& Hartquist (1994), and the effective grain-neutral 
drift speed is set at 1 \(\kms\).  With these choices, the mean grain 
charge \(Z_g = -19.15\), and the ion, electron and grain Hall 
parameters in the pre-shock gas are 4600, \(-2.503\ee 6 \), and \( -0.1606  \)
respectively.

\begin{figure}
\centerline{\epsfysize=7.5cm \epsfbox{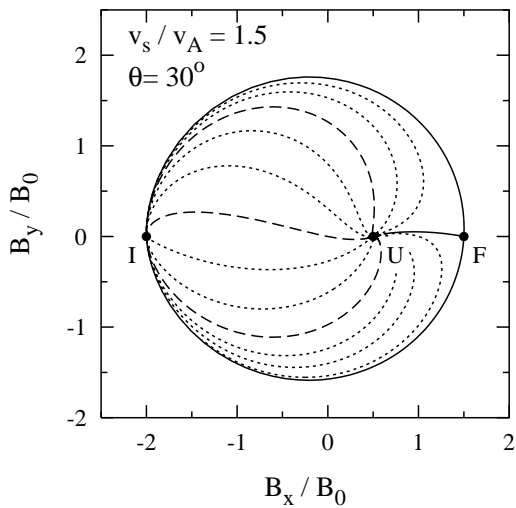}}
\caption{The phase space for shocks with $v_s/v_A =
1.5$ and $\theta=30\degr$.  Each shock solution is represented by a 
trajectory $(B_x(z),B_y(z))$.  The point U indicates the upstream 
state, and the points F and I indicate the downstream states for fast 
and intermediate shocks respectively.  The solid curve running from U 
to F is the trajectory for the fast shock solution, and the two solid 
curves running from F to I are intermediate shocks that can exist in 
the downstream state of the fast shock.  The remaining trajectories 
(dotted and dashed lines) are a one-parameter family of intermediate 
shock waves.  The structure of the three intermediate shock solutions 
indicated by dashed trajectories are plotted in Figures 
\ref{fig-90d-struc}, \ref{fig-270d-struc} and \ref{fig-180d-struc}.}
\label{fig-int-phase-space}
\end{figure}

Both intermediate and fast shocks exist for these parameters, with 
\(B_x/B_0\) equal to 0.5 for the upstream state, and 1.5 and \( -2.0 
\) for the downstream states of fast and intermediate shocks 
respectively.  Before considering the structure of individual shock 
waves it is instructive to examine the entire set of shock transitions 
in the \(B_x\)--\(B_y\) phase space (see \S 2.3).  This is presented 
in Figure \ref{fig-int-phase-space}, with points marked on the \(B_x\) 
axis corresponding to the upstream state (U) and to the downstream 
states for fast (F) and intermediate shocks (I).  Only integral curves 
corresponding to shock solutions, i.e.  beginning and ending at 
stationary points, are plotted .

The one-parameter family of solutions running from U to I are 
intermediate shocks, in which \( \B_\perp \) rotates by 
180\degr\ through the shock front.  A unique solution, the fast shock, 
runs from U to F close to the \(B_x\) axis.  In addition, two 
intermediate shocks run from F to I with opposite senses of transverse 
field rotation.  There are no other acceptable solutions, as the 
remaining curve terminating at F, and the remaining set of curves 
terminating at I are physically unacceptable, with \(\B\) diverging at 
\( z=-\infty \).

The slight asymmetry between the half-planes \(B_y>0\) and \(B_y<0\) 
reflects the asymmetry in the properties of charged particles of 
different sign, namely that the negatively-charged grains are 
partially decoupled from the magnetic field, whereas the remaining 
charged species are well coupled.  If all charged species were 
well coupled, the diagram would be symmetric upon reflection about the 
\(B_x\) axis.
\begin{figure}
\centerline{\epsfxsize=8cm \epsfbox{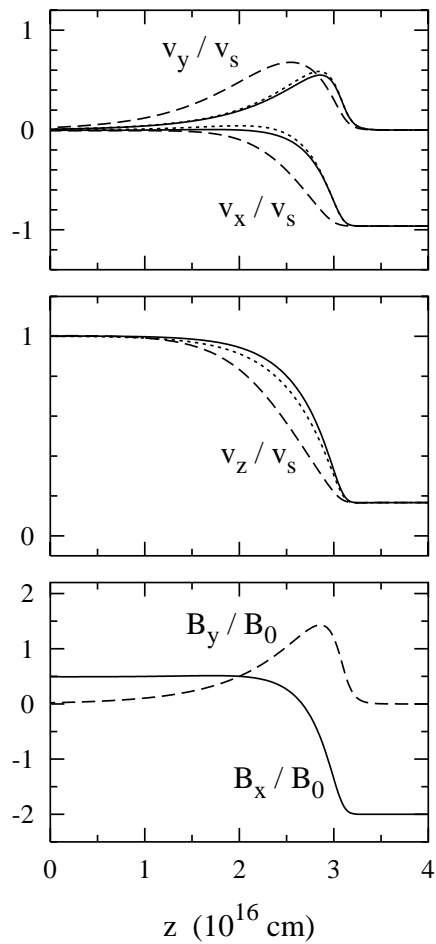}}\vskip 0.2cm
\caption{The structure of the intermediate shock corresponding to the 
dashed trajectory leaving the point U in figure 
\ref{fig-int-phase-space} towards the upper half-plane of phase 
space.  The top panel shows the 
transverse velocity components of the neutrals (solid), ions and 
electrons (dashed), and grains (dotted).  The $z$-component of 
velocity is plotted in the middle panel, and the transverse magnetic 
field components are plotted in the lower panel.}
\label{fig-90d-struc}
\end{figure}
\begin{figure}
\centerline{\epsfxsize=8cm \epsfbox{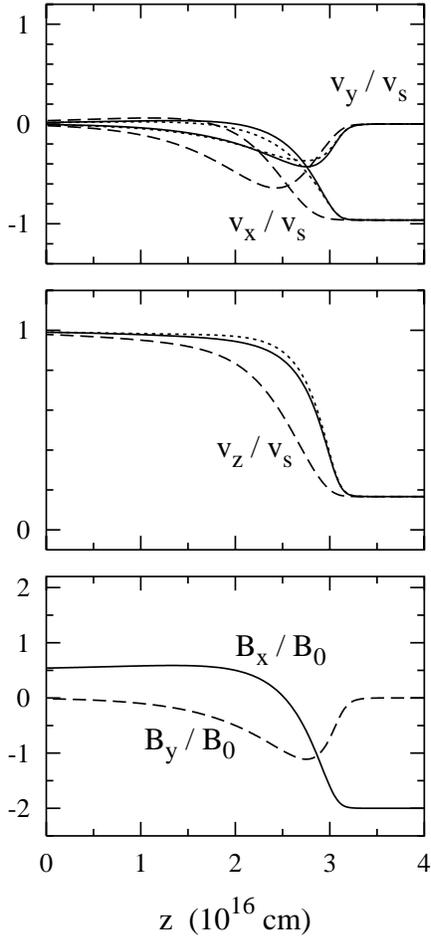}}\vskip 0.2cm
\caption{As for Figure \ref{fig-90d-struc}, but for the intermediate 
shock whose dashed trajectory leaves U in Figure \ref{fig-int-phase-space} towards the 
\emph{lower} half-plane of the phase space.}
\label{fig-270d-struc}
\end{figure}

Figure \ref{fig-90d-struc} and \ref{fig-270d-struc} show the detailed 
structure of two of the intermediate shocks whose trajectories in Fig.  
\ref{fig-int-phase-space} are indicated by dashed lines.  The rotation 
of \( \B_\perp \) has the opposite sense in each shock.  The shocks 
are similar to those presented by Pilipp \& Hartquist (1994) (see 
their Figs 1a\( ' \) --1c\( ' \) and 2a\( ' \)), with \(B_x\) changing 
sign across the shock front, and the \(y\)-components of velocity and 
magnetic field becoming comparable to the \(x\)-component within the 
shock front.  The difference between the shock structures lies 
primarily in the signs of the \(y\)-components.  This would be the 
only difference if all of the charged species were tied to the field 
lines, when the phase space is symmetric under reflection in the 
\(x\)-axis.  However, as the grains are poorly coupled to the magnetic 
field the grain drift through the neutrals is much smaller than 
that of the ions, electrons and magnetic field, and the direction of 
this drift shows distinct differences between the two shocks.  For 
example, the grains are declerated before the neutrals in the middle 
panel of Fig.  \ref{fig-90d-struc}, and lag the neutrals in the middle 
panel of Fig.  \ref{fig-270d-struc}.  Similar differences exist in the 
\(x\) and \(y\) velocity components.

\begin{figure}
\centerline{\epsfxsize=8cm \epsfbox{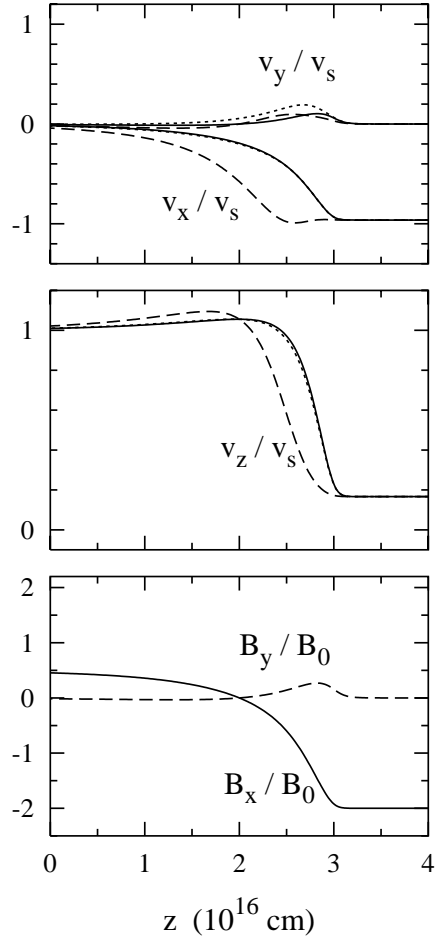}}\vskip 0.2cm
\caption{As for Figure \ref{fig-90d-struc}, but for the 
intermediate shock whose dashed trajectory runs through the origin of 
the phase space of Figure \ref{fig-int-phase-space}.  Note the 
rarefaction precursor ($ z < 2 \ee 16 \u cm $).}
\label{fig-180d-struc}
\end{figure}

Although these two shocks are fairly typical members of the 
intermediate family, shocks with trajectories passing close to the 
origin of phase space cannot be characterised by rotation of 
\(\B_{\perp}\).  The structure of the intermediate shock which has a 
trajectory passing through the origin of phase space (see Figure 
\ref{fig-int-phase-space}) is plotted in Figure \ref{fig-180d-struc}.  
The transverse field does not increase monotonically through the shock 
front, but drops initially.  As the sum of the transverse magnetic 
field \(B_\perp^{2} / 8 \pi\) and the ram pressure \(\rho_0 v_s v_z\) 
is constant through the shock front, there is a compensating increase 
in the \(z\) components of the fluid velocities, and a corresponding drop in 
density until compression begins in the shock (at \(z \ga 2 \ee 16 
\u cm \)).  The shock has a rarefaction precursor, which sucks the 
upstream material towards the shock before accelerating and 
compressing it.

\begin{figure}
\centerline{\epsfxsize=8cm \epsfbox{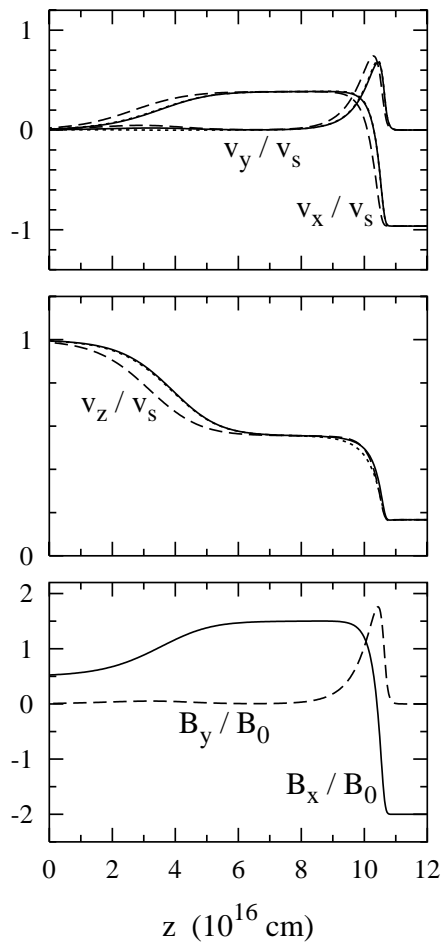}}\vskip 0cm
\caption{As for Fig.  \ref{fig-90d-struc}, but for an intermediate 
shock whose trajectory almost reaches the point F in Fig.  
\ref{fig-int-phase-space} before turning towards I.  The transition 
from upstream to downstream is made in two distinct stages, a fast 
shock transition ($ z<6\ee 16 \u cm $) followed by an intermediate 
shock ($ z>8\ee 16 \u cm $) (see text).} 
\label{fig-fast-int-struc}
\end{figure}

Rather than show the fast shock running from U to F and those running 
from F to I seperately, Figure \ref{fig-fast-int-struc} shows the 
structure of an intermediate shock whose phase-space trajectory lies 
close to the fast shock trajectory, almost reaching F before turning 
and running towards U close to the F to I transition with \(B_y > 0\).  
Because \(dB_x/dz\) and \(dB_y/dz\) become small in the neighbourhood 
of F, the shock structure consists of two distinct parts, 
corresponding to a fast shock (\(z \la 6\ee 16 \u cm \)), followed by 
an intermediate shock (\(z \ga 8\ee 16 \u cm \)).  By choosing 
trajectories that approach successively closer to F before turning 
towards I, the gap between the transitions can be made arbitrarily 
large, becoming infinite in the limit that the trajectory actually 
reaches F. Alternatively, choosing trajectories that turn away at 
successively earlier stages causes the two structures to merge.

The structure of the fast shock is very similar to `standard' 
solutions, with small departures of vector quantities out of the 
\(x\)-\(z\) plane.  This is because the grains are not dominating the 
dynamics of the shock front.  The \(z\)-component of the drift 
velocities is relatively small in the intermediate part of the shock 
transition because the compression across this part of the structure 
is only 4/3, compared to 3 in the fast part of the transition.

\subsection{Strong C-type fast shocks}
\label{subsec-strong-shocks}

\begin{figure}
\centerline{\epsfxsize=8.5cm \epsfbox{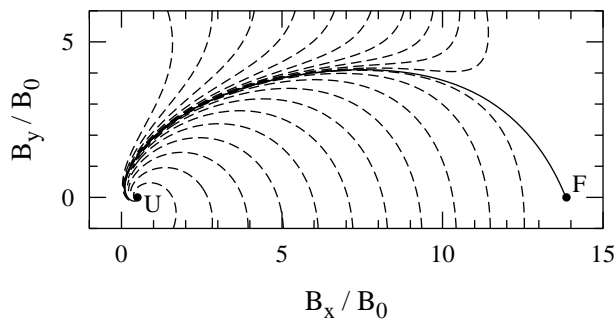}}\vskip 0cm
\caption{Integral curves in the $B_x$-$B_y$ phase space
for $v_s/v_A = 10$, and $\theta=30\degr$. The solid curve running from 
U to F is the fast shock trajectory.  The neighbouring
trajectories (dashed) all correspond to unphysical shock solutions.}
\label{fig-t6-phase-space}
\end{figure}

Now consider stronger shocks with \(v_s/v_A=10\), corresponding to a shock 
speed of 18 \(\kms\).   As these shocks are characterised by larger 
drift speeds and higher temperatures, I adopt 1000 K, 2000 K, and 
10 \(\kms\) as the effective values of the neutral and electron 
temperatures and the grain drift speed.  This implies a mean grain 
charge \(Z_{g}=-191.5\), and ion, electron, and grain Hall parameters of 
4609, \(-7.915\ee 5 \), and \(-0.2420\) respectively.

The shock speed is too high for intermediate shocks to exist as \( 
v_s/v_A > \sqrt{2} \cos \theta \approx 2.45 \).  The fast shock 
running from U to F remains viable.  The structure of the phase space 
around the fast shock trajectory is shown in Figure 
\ref{fig-t6-phase-space}.  Each of the neighbouring trajectories peels 
off and heads towards I, eventually becoming unphysical.  Thus an 
attempted integration of the fast shock solution starting at U will 
stumble onto a trajectory headed for I, and yield only unphysical 
solutions, as found by Pilipp \& Hartquist (1994).  However, 
integration in the reverse direction, from F to U, will be successful.

\begin{figure}
\centerline{\epsfxsize=8cm \epsfbox{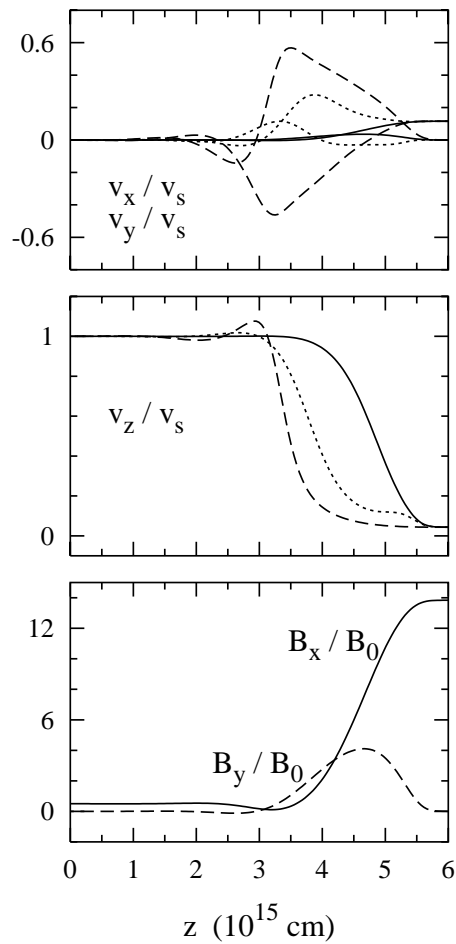}}
\vskip 0cm
\caption{As for Fig. \ref{fig-90d-struc}, but for the fast 
shock of Fig. \ref{fig-t6-phase-space}.  Note 
that the curves in the top panel terminating at 0.1 and 0 are the 
$x$ and $y$-components of the fluid velocities respectively.}
\label{fig-t6-struc}
\end{figure}

The resulting fast shock solution, shown in Figure \ref{fig-t6-struc}, 
is similar to a `standard' C-type shock, apart from departures of 
the magnetic field and fluid velocities out of the \(x\)--\(z\) plane 
at the level of 50 per cent.  The grains lag the ions in the \(x\) and 
\(z\) directions, where they are caught between the magnetic field and 
neutral fluid; and they lead in the \(y\)-direction in which they are 
responsible for the excursion.

\begin{figure}
\centerline{\epsfxsize=8cm \epsfbox{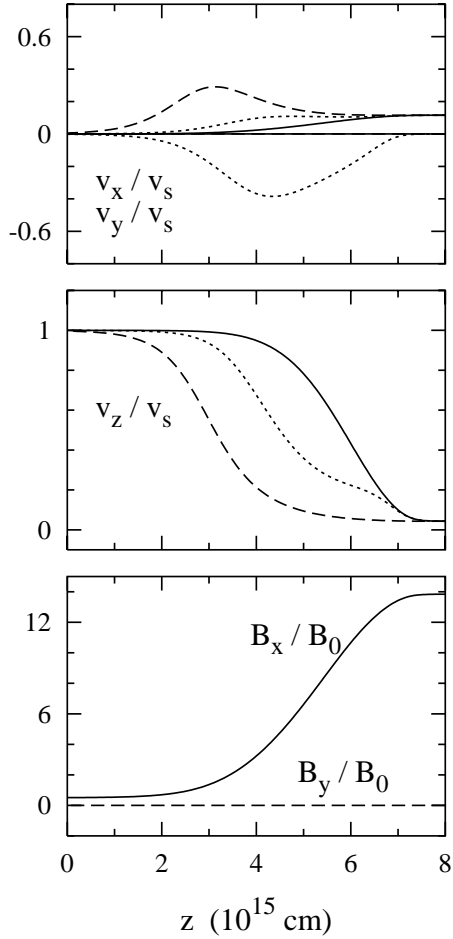}}
\vskip 0cm
\caption{As for Fig. \ref{fig-t6-struc}, but with $ dB_y/dz $ set 
to zero.}
\label{fig-t6-no-y-struc}
\end{figure}
To evaluate the effect of the out-of-plane forces on the shock 
structure, the result of supressing \(B_y\) is plotted in Figure 
\ref{fig-t6-no-y-struc}.  The \(y\)-component of the ion and electron 
velocities are very small, as these species are tied to the field 
lines.  The grains, however still have a substantial drift in this 
direction, but this component of the drag force has been neglected.  
The shock is substantially thicker (\(5\ee 15 
\u cm \)) than the full solution shown in Figure \ref{fig-t6-struc} 
(\(3\ee 15 \u cm \)), 
and drift velocities are much higher in the full solution as the 
kinetic energy flux that must be dissipated into thermal energy is the 
same in both cases.  The temperature within the shock of Fig.  
\ref{fig-t6-struc} will be significantly higher.  Although the 
temperature cannot be calculated in our prescription, it is useful to 
examine the heating rate per unit mass associated with the collisions 
between neutrals and charged species,
\begin{equation}
\frac{G}{\rho} = \sum_{j}^{} \gamma_j \rho_j |\v_j - \v|^2
	\label{eq-heating}
\end{equation}
(Draine 1986; Chernoff 1987), which would largely be
balanced by the local radiative 
cooling rate per unit mass.  \( \gamma_j \rho_{j0} \) is \( 1.485\ee 
-11 \), \( 8.649\ee -14 \), \( 1.354\ee -10 \ut s -1 \) for the ions, 
electrons and grains respectively.  Thus for a given drift speed, the 
grains contribute a drag force and collisional heating rate ten times 
that of the ions, and the heating by electron collisions is negligible.
\begin{figure}
\centerline{\epsfxsize=8cm \epsfbox{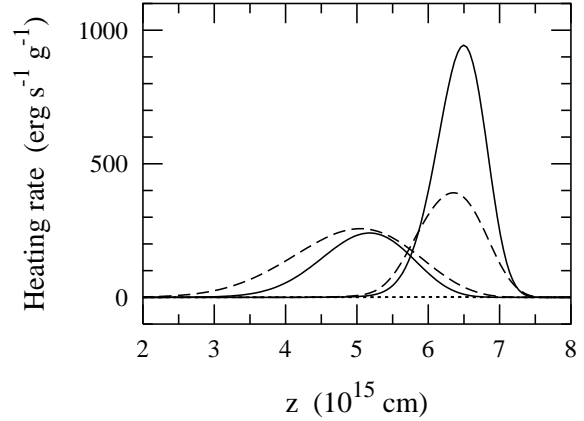}}
\vskip 0cm
\caption{The heating rate per unit mass within the shock fronts of 
figures \ref{fig-t6-struc} (peaks at $ z\approx 5\ee 15 \u cm $) and 
\ref{fig-t6-no-y-struc} (peaks at $ z\approx 6.5\ee 15 \u cm $).  
For each shock the heating associated with grain-neutral, ion-neutral, 
and electron-neutral collisions are plotted as solid, dashed, and 
dotted curves respectively.}
\label{fig-t6-heating}
\end{figure}
The actual heating rate for the different species within the two shocks is 
plotted in Figure \ref{fig-t6-heating}.  As expected, the heating is 
more severe in the full solution.  However, note that most of the 
difference is due to the grain collisions which \emph{dominate} the 
ion collisions in the full solution.  It is clear, therefore, that the 
artificially coplanar solutions that have been used in detailed 
C-shock models (Draine et al. 1983; Kaufman \& Neufeld 
1996a,b) may substantially understimate the
the peak shock temperature.

Now consider the effect of raising the pre-shock density to 
\(10^{7}\percc \).  This increases the importance of the grains, as 
the pre-shock ion fraction is decreased by a factor of three, and the 
grains are further decoupled from the magnetic field.  The pre-shock 
ion, electron, and grain Hall parameters in this case are 460.9, 
\(-7.915\ee 4 \), and \(-0.02420\) respectively.
\begin{figure}
\centerline{\epsfxsize=8cm \epsfbox{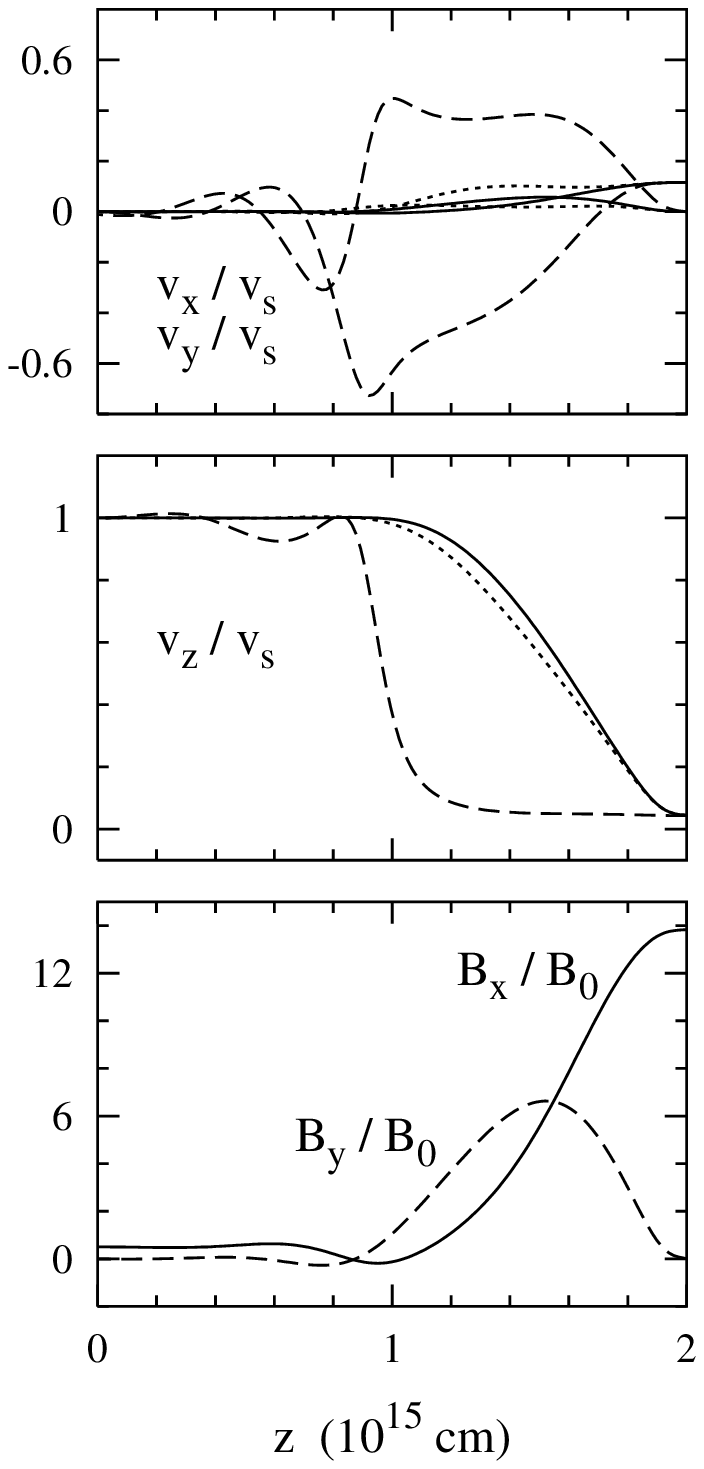}}
\vskip 0cm
\caption{As for Fig. \ref{fig-t6-struc},
but with pre-shock density $10^{7} \percc $.}
\label{fig-t18-struc}
\end{figure}
The solution is plotted in Figure \ref{fig-t18-struc}.  As expected, 
the \( y \)-components of the fluid velocities and magnetic field are 
increased, and the grain drift through the neutrals is substantially 
decreased.  There is also the the beginnings of oscillatory behaviour 
in the shock precursor, corresponding to the spiralling of the shock 
trajectory out of the point U in the shock plane (see Fig.  
\ref{fig-fast-phase-space}).

\begin{figure}
\centerline{\epsfxsize=8cm \epsfbox{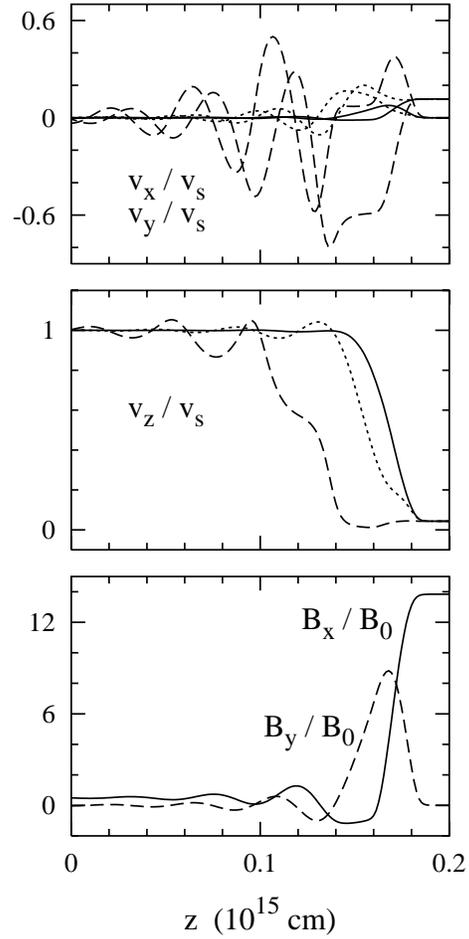}}
\vskip 0cm
\caption{As for Fig. \ref{fig-t6-struc},
but with pre-shock density $10^{7} \percc $ and 0.1 $\mu$ grains.}
\label{fig-t13-struc}
\end{figure}

Finally, consider the effect of decreasing the grain size from 0.4 to 
\(0.1 \mu\), which increases the grain abundance by a factor of 64 to 
\(n_g/\nh = 1.6\ee -12 \).  As the mean grain charge is reduced by a 
factor of 4, to \(Z_g=-47.87\), this increases the charge resident on 
grains by a factor of 16, and grains now dominate the current within 
the shock front.  The shock solution is plotted in Fig.  
\ref{fig-t13-struc}.  Although the ion-neutral streaming velocities 
are a significant fraction of the shock speed in the precursor, most 
of the energy is dissipated by grain-neutral collisions within the 
extremely thin main part of the shock front (\( z=0.14 \)--\(0.18 \ee 
15 \u cm \)) -- ions would have to stream through the neutrals over a 
length scale of order \( 5\ee 15 \u cm \) to dissipate the incoming 
kinetic energy.  Besides producing significant partial rotation of the 
field and fluid velocities within the shock front, the precursor of 
the shock exhibits an oscillatory behaviour because the propagation of 
wave modes in the pre-shock gas at these length scales is substantially 
modified.  This is manifested in the linear analysis near the 
stationary points, which shows that the upstream state becomes a 
strongly spiral node.
\begin{figure}
\centerline{\epsfxsize=9cm \epsfbox{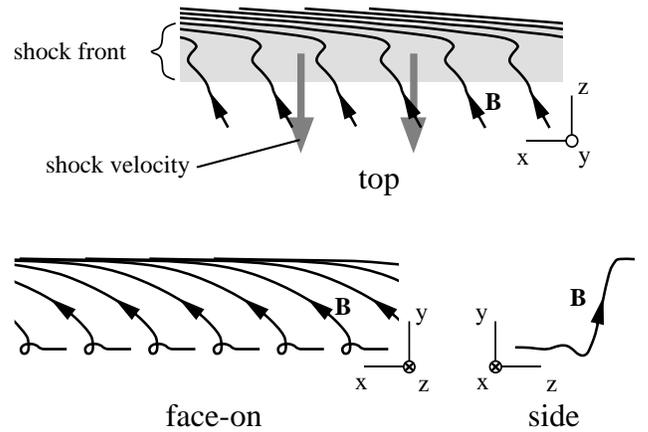}}
\hskip -2cm
\vskip 0cm
\caption{The magnetic field structure in the shock solution plotted in
Fig. \ref{fig-t13-struc}.}
\label{fig-shock-sketch}

\end{figure}
The ion and electron velocities approach zero and their densities 
rapidly increase at \(z\approx 0.15 \ee 15 \u cm \), at which point 
large currents prevent further movement to the left in the 
\(B_x\)--\(B_y\) phase space (see Fig.  \ref{fig-fast-phase-space}) 
and the \(B_{x}\) and \(v_x\) profiles are clipped.  The field-line 
structure of the shock is sketched in Fig. \ref{fig-shock-sketch}.

\begin{figure}
\centerline{\epsfxsize=8.5cm \epsfbox{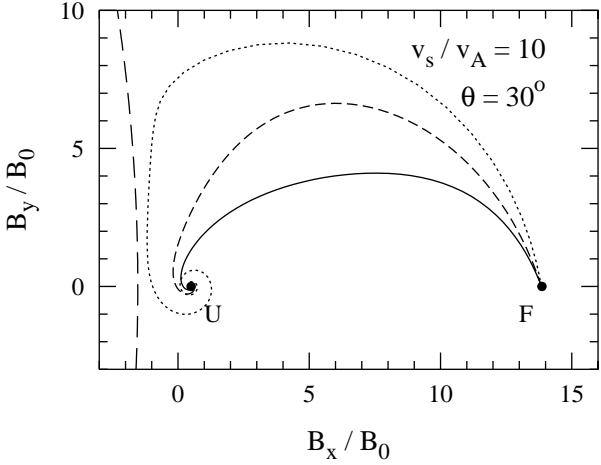}}
\vskip 0mm
\caption{The phase portraits for the fast shock solutions of Figures 
\ref{fig-t6-struc} (solid), \ref{fig-t18-struc} (short-dashed) and 
\ref{fig-t13-struc} (dotted). 
The long-dashed curve indicates where the ion and electron 
densities diverge (see text).}
\label{fig-fast-phase-space}
\end{figure}
The trajectories of the shock solutions are plotted in Figure 
\ref{fig-fast-phase-space}.  As expected, the shocks are no longer 
close to being coplanar, and the upstream state becomes a spiral node, 
with the magnetic field oscillating ahead of the shock solution.  The 
long-dashed curve shows the locus where the the ion and electron 
densities diverge, i.e.  where \( v_{iz}=v_{ez}=0 \) (see \S 
\ref{subsec-forbidden}).  The trajectory for the shock 
propagating into the \( 10^7 \percc \) gas with 0.1 \( \mu \) grains 
approaches this locus, but is forced away by the associated large 
current density.

\subsection{The effect of the grain-size distribution}
\label{subsec-grain-size-results}

In this subsection, I consider the effect of an MRN distribution on 
the shock dynamics, both with and without an additional population of 
very small grains.  Ionization models of the ambient molecular gas 
show that at gas densities of order \( 10^6 \percc \), not all grains 
are charged, but the no.  density of those that are is order \( 
10^{-11} \nh \).  Thus, the net charge on the grains, \( n_g Z_g 
\approx 10^{-11} \nh\) is of the same order as that for the 
single-grain-size models presented in the previous subsection, for 
which \( n_g / \nh \approx 10^{-13} \) and \( Z_g \approx -120 \).  
The shock models are therefore rather similar.

When PAHs are present, they are the dominant charged species.  The 
effects of interest are greatest at densities around \( 10^6 \percc 
\), as at higher densities the abundances of positively and negatively 
charged grains become very similar, and \( |\sigma_1|/\sigma_2 \) 
becomes small.  For \( \nh=10^6\percc \), the fractional abundances of 
positively charged PAHs plus ions is \( 3\ee -9 \), with a similar 
abundance of negatively charged PAHs.  Singly-charged grains and 
electrons both have abundances of approximately \( 2\ee -11 \) (Nishi 
et al. 1991; Kaufman \& Neufeld 1996a).  Thus I adopt 
pre-shock densities and magnetic fields \( \nh=10^6 \percc \) and \( 
B_0 = 1 \u mG \) and a four fluid model consisting of neutrals, 
positively and negatively charged ion-like particles each with \( 
n/\nh=2\ee -11 \), negatively charged grains with \( N/\nh = 2\ee -11 
\).  The two species of ion-like particles are assumed to be identical 
apart from the sign of their charge, with \( <\sigma v> = 1.6\ee -9 
\ut cm -3 \ut s -1 \), so their pre-shock Hall parameters are \( \pm 
4609 \).  The grains are assumed to satisfy the power-law size 
distribution (\ref{eq-mrn}) between \( a_1=50 \)\AA\ and \( a_2=2500 
 \)\AA.  Given a choice of the pre-shock Hall parameter of the smallest 
grains, \( \beta_1 \), the effective values of \( n_g \) and \( 
\beta_g \) can be calculated.
\begin{figure}
\centerline{\epsfxsize=8.5cm \epsfbox{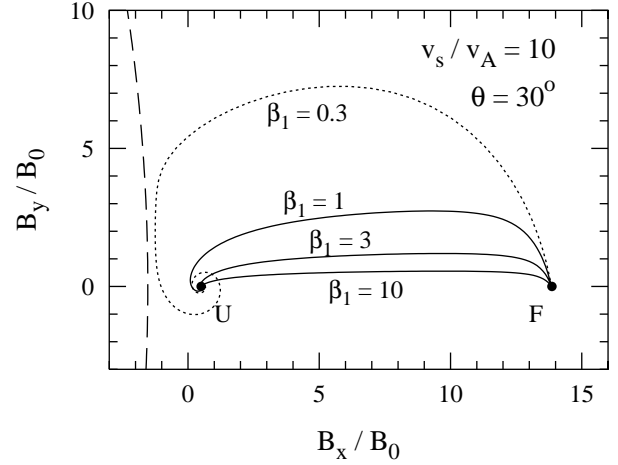}}\vskip 0cm
\caption{Integral curves in the $B_x$-$B_y$ phase space for shocks 
with $ \nh=10^6 \percc $, $v_s/v_A = 10$, and $\theta=30\degr$, for 
the case when grain sizes follow an MRN distribution and there is an 
additional population of PAHs.  The different 
trajectories are labelled by the value of the pre-shock Hall parameter 
for the smallest grains, with radii of 50 \AA\ (see text).} 
\label{fig-PAH-phase-space}
\end{figure}
Rather than show the detailed shock structure, which is similar to 
that presented previously, I show in Figure \ref{fig-PAH-phase-space} 
the phase-space trajectories for different choices of \( \beta_1 \).  
As \( \beta_1 \) decreases, \( \beta_g \) decreases, and \( n_g \) 
increases, so the shocks become increasingly non-coplanar, with the 
characteristic spiral trajectory for \( |\beta_g| \la 0.3 \).  This is 
to be expected from the non-coplanarity criterion 
(\ref{eq-noncoplanar-criterion}).  These results indicate that 
significant non-coplanarity occurs for \( \beta_1 \la 1 \), that is, 
in regions of unusually weak field (see eq [\ref{eq-beta1}]).

In the absence of PAHs, the dominant charged fluids for \( \nh \la 
10^9 \percc \) are ions, electrons and grains carrying a single 
negative charge (Nishi et al. 1991).  At densities of \( 
10^6,\,10^7,\, 10^8 \percc \), I adopt fractional abundances \( 
n_i /\nh=10^{-9},\, 3\ee -10 ,\, 1.03\ee -10 \) and \( N/\nh=3\ee 
-10 ,\, 2\ee -10 ,\, 1\ee -10 \) respectively.  At the two lower 
densities, the density of electrons is determined by charge 
neutrality once the charged grain abundance is corrected for the size 
distribution.  At \( 10^8 \percc \), the electron density is 
relatively small and would be strongly affected  by this procedure, 
so the no. density for charged grains is not corrected for the size 
distribution.  In any case, 
the corrections would amount to changes of order 5 per cent.
\begin{figure}
\centerline{\epsfxsize=8.5cm \epsfbox{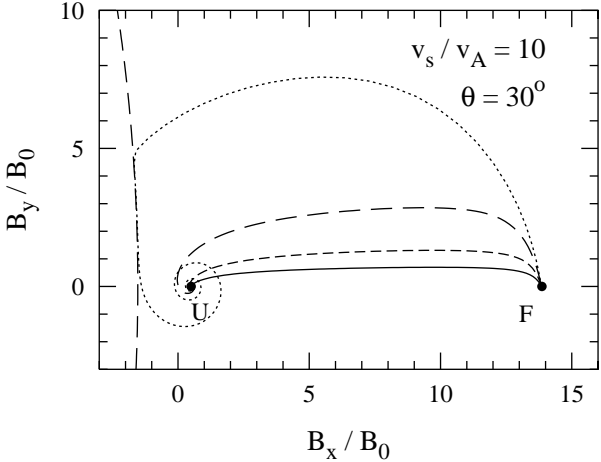}}\vskip 0cm
\caption{Integral curves in the $B_x$-$B_y$ phase space for shocks 
with $v_s/v_A = 10$, and $\theta=30\degr$, for 
the case when grain sizes follow an MRN distribution and PAHs are 
absent.  The different 
trajectories correspond to different choices of pre-shock magnetic field 
and gas density -- \emph{solid:}$ 1 \u mG,\,10^6 \percc $;
\emph{short-dashed:}$ 0.3 \u mG,\,10^6 \percc $ or $ 3 \u mG,\,10^7 \percc 
$;  \emph{long-dashed:} $ 1 
\u mG,\,10^7 \percc $ or $ 10 \u mG,\,10^8 \percc $;  \emph{dotted:} $ 3 
\u mG, \,\,10^8 \percc $.} 
\label{fig-noPAH-phase-space}
\end{figure}
Figure \ref{fig-noPAH-phase-space} presents the trajectories for 
shocks with \( v_s/v_A =10 \) at these three densities for standard 
and lower values of the magnetic field.  This illustrates how the 
shock trajectory depends primarily on the ratio \( B/\nh \) that 
determines the Hall parameters, with the particle abundances and field 
strength determining the shock thickness.

\section{Discussion}
\label{sec:discussion}

The results presented in \S\ref{sec-results} confirm that both intermediate and fast 
C-type shock solutions exist.  A one-parameter family of intermediate 
shocks exists for each shock speed between the intermediate speed \( 
v_A \cos\theta \) and \( \sqrt{2} v_A \cot \theta \); at higher speeds 
the downstream states become unphysical.  A unique fast shock exists 
for each shock speed in excess of the fast speed \( v_{A} \).  The 
relationship between these two classes of solution is summarised by 
the trajectories in the \( B_{x} \)--\( B_{y} \) phase-space diagram.

The fast shock trajectory runs along the \( B_{x} \) axis if the 
charged species are tied to the magnetic field lines, or if positive 
and negatively charged species have identical abundances and Hall 
parameters.  In this case, integration of the fast shock solution away 
from the upstream state is straightforward.  The direction of the 
trajectory leaving the upstream stationary point (i.e.  along the 
\(B_x\) axis) is known, and the velocity and magnetic field vectors in 
the fast shock solution are confined to the \(x\)-\(z\) plane, so the 
\( y \) components can be discarded.  Integration can be started by 
stepping off U towards F, and then integrating the trajectory (e.g.  
Draine 1980; Wardle \& Draine 1987).

The symmetry of phase space under reflection in the \( B_x \) axis is 
destroyed by grains, which being partially decoupled from the magnetic 
field and preferentially carrying a negative charge, introduce a 
handedness into the fluid.  The fast shock trajectory no longer runs 
along the \( B_{x} \) axis, and this distortion provides an 
opportunity for numerical mixing of the fast solution with the 
neighbouring divergent intermediate shock trajectories during 
integration.  Any attempted integration from U to F is doomed.  To 
begin with, there is no {\it a-priori} means of determining the 
direction for the initial step off the point U. In principle this 
problem could be circumvented by using a shooting method to adjust the 
starting direction until the integration successfully reached F. 
However, because of finite numerical precision, any attempted 
integration into F along the fast trajectory will step onto a 
neighbouring intermediate shock trajectory and eventually be forced 
away from the fast trajectory and towards I (c.f.  integration towards 
sonic points).  This behaviour prevented Pilipp \& Hartquist (1994), 
who started their integrations at U and treated the direction of the 
first step away from U as a free parameter, from finding the fast 
shock solutions.  At low shock speeds, they discovered the family of 
intermediate shocks but could not find the fast shock, noting instead 
that the sense of rotation of \( \Bp \) within the intermediate shocks 
changed at a critical direction of the initial step away from U. At 
higher speeds, for which the intermediate shock solutions become 
unphysical, Pilipp \& Hartquist were unable to find any acceptable solutions.

A realistic treatment of the ionization balance is required for a 
quantitative assessment of the role of grains in C-type shocks.  The 
ionization balance is particularly important when \( v_z \) for any 
charged species approaches zero and the number density of that species 
diverges in the constant-flux approximation (eq. \ref{eq-jcontinuity}) 
used here.  This does not affect the divergence of trajectories from 
the fast trajectory unless the fast trajectory itself approaches this 
limit (e.g.  the trajectory shown in Fig. 
\ref{fig-fast-phase-space} for the shock displayed in Fig.  
\ref{fig-t13-struc}).  

Some conclusions are robust to changes in the microphysics.  If the 
conductivity of the gas can be regarded as some unspecified function 
of local physical variables at any point within the shock front, so 
that shock solutions can still be represented as trajectories in the 
\( B_y \)--\( B_x \) phase-space, the topology of the phase space is 
determined by the nature of the fixed points.  These were shown in \S 
\ref{subsec-classification} to be a source, sink and saddle 
independent of the conductivity tensor.  Thus the integral curves 
always have the topology plotted in Figure \ref{fig-int-phase-space}, 
apart from smooth distortions (as apparent in Fig 
\ref{fig-t6-phase-space}) because of changes to the conductivity 
tensor.  The intermediate-shock curves always run from U or D to I, 
and thus \( v_{iz}=v_{ez}=0 \) at some point if the shock speed is 
high enough for intermediate shocks to be excluded.  The fast shock 
trajectory could in principle cross the \( v_{iz}=v_{ez}=0 \) locus if 
the ionization balance is able to distort the trajectory sufficiently, 
although it is apparently unable to do so in the fixed-flux 
approximation used here (see Fig.  \ref{fig-noPAH-phase-space}).  In 
any case, when the fixed-flux condition is dropped, crossing this 
locus does not imply infinite charged-particle abundances as the 
increase is limited by recombinations.  Intermediate shocks, however, 
will still be excluded as their downstream conditions are unphysical.

A more realistic treatment of the physical processes occurring within 
the shock front is essential as the grain properties within the shock 
and the shock structure itself are intimately related.  In particular, 
the neutral and electron temperatures are required to accurately 
calculate the Hall parameters for a given grain size.  In addition, 
the grain-size distribution varies within the shock front as different 
sized grains will generally have different drift speeds, the smallest 
grains being compressed with the magnetic field, the largest grains 
with the neutrals.  Incorporation of these effects increases the 
dimensionality of the phase space by introducing differential 
equations for the temperatures and relative abundances of different 
species.  In principle this could alter the topology of the phase 
space, although this is unlikely to be the case in practice.  There 
are, however, practical problems in finding the fast shock structure 
by integrating backwards from the downstream state if chemistry is 
coupled to the hydrodynamics (through e.g.  changing the abundances of 
coolants or the degree of ionization in the shock), since one does not 
then know the post-shock conditions {\it a-priori} (c.f.  the 
discussion in Roberge \& Draine (1990) regarding C\(^*\)- and 
J-shocks), and a time-consuming shooting method must be employed.

Although the grain drag significantly modifies the shock structure, it 
is unclear whether this will produce gross differences in the 
total line emission from the shock front.  The essential feature of 
C-shock structure is that it maintains a layer of hot molecular gas.  
For a given pre-shock density and shock speed, the same power is 
dissipated per unit area independently of the detailed structure of 
the shock front.  The gas generally will be heated up to temperatures 
between 1000 and 2000 K at which point the local heating and cooling 
balance.  Molecules will remain intact and the molecular line emission 
will not be greatly affected.  The insensitivity of line ratios to 
details of shock structure is illustrated by recent numerical work 
(Stone 1997; Neufeld \& Stone 1997; MacLow \& Smith 1997), which 
follows the development and saturation of the instability in C-type 
shocks (Wardle 1990).  Even though the neutral gas is collected into 
dense fingers within the shock front, the line ratios are generally 
unchanged by more than a factor of two (Neufeld \& Stone 1997), 
although diagnostics can be found (MacLow \& Smith 1997). 
Nevertheless, one might expect that the grains will affect the 
stability of C-shocks and also the speed at which shocks become J-type.

Finally, it is worth pointing out that the theoretical study of 
C-shocks in molecular clouds will contribute to the general theory of 
MHD shock waves, and of intermediate shocks in particular (see the 
review by Wu 1995).  Until recently intermediate shocks in MHD were 
thought to be unphysical because they appeared to lack sufficient 
freedom to adjust to slight perturbations in the upstream or 
downstream flow (see, e.g.  Kantrowitz \& Petschek 1965).  The issue 
was reopened when intermediate shocks emerged as stable structures in 
numerical simulations (Wu 1988a; Steinolfson \& Hundhausen 
1990a,b).  It has since been realised that the arguments against the 
existence of intermediate shocks are based on ideal MHD, which breaks 
down completely within the shock front.  Numerical studies of 
time-dependent intermediate shocks in {\it resistive} MHD demonstrated 
that the shocks could be formed `naturally' and were stable (Wu 
1990).  An examination of the structure of resistive, intermediate 
shocks in the weak shock limit (Kennel et al. 1989; Wu \& 
Kennel 1992), shows that, as found for C-type shock waves, for given 
external parameters (i.e.  pre-shock conditions and shock speed), a 
one-parameter family of intermediate shock structures exist connecting 
the jump conditions.  This `internal' parameter provides the 
additional freedom for the shock to adjust its structure to external 
perturbations, a behaviour outside of the scope of ideal MHD in which 
the shock transition is represented as a discontinuity between the 
upstream and downstream states.  As these conclusions depend on shock 
structure, in principle they depend on the model used for the 
magnetised medium.  Resistive (Wu 1990) and hybrid (Wu \& Hada 1991) 
models have been investigated so far.  With the advent of 
multidimensional ambipolar diffusion codes, it has become possible to 
study these issues in weakly ionized media (see Smith \& MacLow 1997).

\section{Summary}
\label{sec:summary}

In this paper I constructed steady models of oblique C-type shocks in 
which the magnetic field within the shock front is not artificially 
confined to the plane containing the upstream and downstream magnetic 
field and the shock normal.  Four fluids -- neutrals, ions, electrons 
or PAHs, and negatively charged grains -- were considered.  The 
thermal pressure of the fluids and the inertia of the charged 
components were neglected.  The effects of chemistry and changes in 
fractional ionization, and the charge residing on grains were ignored.  
The rate coefficients for grain-neutral and electron-neutral elastic 
scattering (which depend in principle on the neutral temperature and 
the grain-neutral drift speed, and on the electron temperature, 
respectively) were assumed constant with values appropriate to 
conditions within the shock front.

These assumptions reduce the number of differential equations describing 
the shock structure to a pair for the components \( B_x,\,B_y \) of 
the magnetic field that are perpendicular to the shock normal.  Shock 
solutions can therefore be conveniently represented by trajectories in 
a two-dimensional phase space.

Models were presented for weak shocks (\( v_s/v_A = 1.5 \)), and for 
strong shocks (\( v_s/v_A =10 \)).  Variations in the grain size from \( 
0.1 \mu \) to \( 0.4 \mu \), and the effect of an MRN grain-size 
distribution were considered.

The results are summarized as follows:

\begin{enumerate}
\item
The cold MHD jump conditions permit three stationary points in the 
phase space, corresponding to the upstream state and downstream states 
of the fast (for \(v_s > v_A\)) and intermediate shocks (\(v_A\,{\rm 
cos}\,\theta < v_s < v_A {\rm cot}\,\theta\)).  Valid 
shock solutions have phase-space trajectories linking the upstream 
state to one of the downstream states.  A linear analysis, valid for 
\emph{any} number of charged species, shows that the upstream state is 
a source, the fast downstream state is a saddle, and the intermediate 
downstream state is a sink. 
\item  
A one-parameter family of intermediate shocks exists for each shock 
speed in the range \(v_A\,{\rm cos}\,\theta < v_s < v_A {\rm 
cot}\,\theta\) where \(\theta\) is the angle between the shock normal and 
the pre-shock magnetic field.  These solutions correspond to those 
found by Pilipp \& Hartquist (1994).  The family contains members with 
either sense of rotation of \( \Bp \) through the shock front, members 
with a rarefaction precursor in which \( \Bp \) becomes small within 
the front before compression begins, and members that correspond to a 
fast shock followed downstream at a finite distance by an 
intermediate shock.
\item  
A unique fast shock exists for each shock speed \(v_s > v_A\), where 
\(v_A\) is the Alfv\'en speed in the pre-shock gas. Pilipp \& Hartquist (1994) 
were unable to find 
these solutions because integration of the equations for shock 
structure from the upstream state to the downstream state is unstable.  
Instead, integration must begin at the downstream state and run 
backwards through the shock front.
\item 
When all charged particles are well couwell coupled to the magnetic field, or 
when there is a symmetry between the poorly-coupled particles of 
either sign, the phase space is symmetric about the \( B_x \) axis, 
and the fast shock is coplanar, its trajectory running along the \( 
B_x \) axis.  In conditions typical of molecular clouds, 
negatively-charged grains contribute significantly to the drag and are 
loosely coupled to the field.  The net asymmetry in the coupling of 
negatively and positively charged particles to the magnetic field 
lines imposes a handedness on the shock structure and the phase space 
trajectories lose their symmetry on reflection in the \( B_x \) axis.  
In particular, the fast shock trajectory no longer runs along the axis 
-- the magnetic field and fluid velocities no longer lie in the 
\(x\)-\(z\) plane (containing the pre-shock field and shock normal) but 
have significant \(y\) components within the shock front.
\item 
For typical conditions in molecular clouds,  grains dominate the frictional 
force and the magnetic field is well-tied to ions, electrons and 
PAHs.  Under these conditions for given values of pre-shock magnetic 
field and gas density, the pre-shock grain Hall parameter 
determines the shock structure apart from an overall length scale for 
the shock thickness which is determined by the grain abundance.  Thus 
under the approximations adopted in this paper, the shock structures 
are relatively insensitive to the details of the grain-size 
distribution and the exact composition of the charged species in the 
gas.  
\item
An MRN grain-size distribution can be approximately incorporated by 
calculating an effective abundance and an effective grain Hall 
parameter for single-size grains.  The ratio of the
effective abundance to 
the abundance of charged MRN grains, and the effective Hall 
parameter are determined by the Hall parameter of the smallest grains 
in the MRN distribution.
\item
The degree of noncoplanarity is determined by the grain Hall parameter 
\( \beta_g \), with significant effects being found for \( |\beta_g| \la 
1\).  When \( |\beta_g| \la 0.3 \) the upstream state becomes a spiral 
node and the shocks exhibit a precursor in which \( \Bp \) may make 
several rotations.
\item
Ions and electrons become passive in the sense that they may 
stream through the neutral gas with large transverse drift velocities 
(of order half the shock speed or more) without 
generating significant dissipation, as the shock thickness is 
determined by the collisions of neutrals with charged grains.  The 
primary role of these charged species is to guarantee that the electric 
field in the shock front is very nearly orthogonal to the magnetic field.
\item
Supressing the out-of-plane components of the drift velocities and 
magnetic field may significantly reduce the current and leads to a 
thicker shock structure (by, e.g. a factor of two) and a consequent 
decrease in the heating rate within the shock front.  Models 
supressing the out-of-plane components of the magnetic field and 
velocity therefore underestimate the temperatures within the shock 
front, and significantly underestimate the magnitude of the grain 
drift speed through the neutrals.
\end{enumerate}

This work was initiated at the University of Rochester.  
A. Perez-Miller is thanked for assistance with coding.  This 
research was partially supported by NASA to the University of 
Rochester through grant NAGW-2444.  The Special Research Centre for 
Theoretical Astrophysics is funded by the Australian Research Council 
under the Special Research Centre programme.

\bsp
\label{lastpage}
\end{document}